\documentclass[aps,prb,onecolumn,showpacs,amsmath,amssymb,superscriptaddress,longbibliography]{revtex4-1}

\usepackage{bm}
\usepackage{amsthm}
\usepackage{amsfonts}
\usepackage{graphicx}
\usepackage{epsfig}
\usepackage{subfigure}
\usepackage[colorlinks=true,citecolor=blue,urlcolor=blue]{hyperref}
\usepackage{color}
\usepackage{ulem}

\newcommand{\be}{\begin{equation}}
\newcommand{\ee}{\end{equation}}
\newcommand{\ba}{\begin{eqnarray}}
\newcommand{\ea}{\end{eqnarray}}
\newcommand{\ban}{\begin{eqnarray*}}
\newcommand{\ean}{\end{eqnarray*}}

\newcommand{\one}{\leavevmode\hbox{\small1\normalsize\kern-.33em1}}

\begin{document}

\title{Coherence of an extended central spin model with a coupled spin bath}
\author{Pengcheng Lu}
\affiliation{Institute of Modern Physics, Northwest University, Xi'an 710127, China}
\affiliation{Shaanxi Key Laboratory for Theoretical Physics Frontiers, Xi'an 710127, China}
\author{Hai-Long Shi}
\affiliation{School of Physics, Northwest University, Xi'an 710127, China}
\author{Like Cao}
\affiliation{School of Physics, Northwest University, Xi'an 710127, China}
\author{Xiao-Hui Wang}
\affiliation{Shaanxi Key Laboratory for Theoretical Physics Frontiers, Xi'an 710127, China}
\affiliation{School of Physics, Northwest University, Xi'an 710127, China}
\author{Tao Yang}
\email{yangt@nwu.edu.cn}
\affiliation{Institute of Modern Physics, Northwest University, Xi'an 710127, China}
\affiliation{Shaanxi Key Laboratory for Theoretical Physics Frontiers, Xi'an 710127, China}
\affiliation{School of Physics, Northwest University, Xi'an 710127, China}
\affiliation{Peng Huanwu Center for Fundamental Theory,  Xi'an 710127, China}
\author{Junpeng Cao}
\email{junpengcao@iphy.ac.cn}
\affiliation{Peng Huanwu Center for Fundamental Theory,  Xi'an 710127, China}
\affiliation{School of Physical Science, University of Chinese Academy of Sciences, Beijing 100049, China}
\affiliation{Beijing National Laboratory for Condensed Matter Physics, Institute of Physics, Chinese Academy of Sciences, Beijing 100190, China}
\affiliation{Songshan Lake Materials Laboratory, Dongguan, Guangdong 523808, China}
\author{Wen-Li Yang}
\email{wlyang@nwu.edu.cn}
\affiliation{Institute of Modern Physics, Northwest University, Xi'an 710127, China}
\affiliation{Shaanxi Key Laboratory for Theoretical Physics Frontiers, Xi'an 710127, China}
\affiliation{School of Physics, Northwest University, Xi'an 710127, China}
\affiliation{Peng Huanwu Center for Fundamental Theory,  Xi'an 710127, China}

\date{\today}


\begin{abstract}
We study an extended central spin model with an isotropic nearest-neighbour spin-exchange interaction among the bath spins. The system is controllable by external magnetic fields applied on the central spin and the bath, respectively. We construct a basis set of the Hilbert space and express the Hamiltonian of the extended model into a series of $2\times2$ block matrices to obtain the exact solution of the model successfully. Therefrom, the coherence and the spin polarization of the central spin are investigated. We find that if the couplings among the bath spins are antiferromagnetic, the central spin has good coherence and polarization at low temperatures. Moreover, the decoherence is greatly suppressed at the critical point where the strengthes of the central and the bath magnetic field take the same value. A dephasing phenomenon is identified when the initial state of the central spin is unpolarized.
\end{abstract}

\pacs{75.10.Pq, 02.30.Ik, 71.10.Pm}

\maketitle


\section{Introduction}
\label{sec1} \setcounter{equation}{0}

The central spin model, describing one central spin surrounded by a number of bath spins, was first introduced and solved by Gaudin in 1970s\,\cite{JP.37.1087}. 
In the past decades, the study of central spin systems and the related generalizations has gained great attention because the model has appeared in a large variety of nanostructures, such as semiconductors\,\cite{PRB.76.153302,PRB.76.045218,PSS.246.22032215}, quantum dots\,\cite{PRL.88.186802,PRE.67.056702,PRB.70.195340,PRB.74.195301,PRB.76.014304,PRL.102.057601,PRB.79.245314,PRB.81.165315,PRL.105.177602,PRB.84.155315,PRL.110.040405,PRA.89.012107,PRA.96.052125,JPCM.15.R1809}, carbon nanotubes, and nitrogen vacancy centers in diamond\,\cite{MRS.38.802,PRep.528.1}. This model is closely related to the spin qubit suitable for quantum information processing, quantum computation and quantum metrology devices. The coupling between the central spin and the bath spins induces many interesting phenomena such as coherence\,\cite{JSup.16.221}, many-body localization\,\cite{PRB.98.161122}, the Loschmidt echo\,\cite{NPB.933.454}, and the solution Lee-Yang zero\,\cite{JPA.51.505001}. Great efforts are made to understand the sources and mechanisms of decoherence, the major obstacle for real-life functioning implementations of the model. The exact solutions of central spin systems, giving the many-particle eigenvalues and eigenstates, are important to provide some believable and quantitative results for understanding the dynamical behavior of the system\,\cite{PRB.82.161308}. However, only the systems with an initially fully polarized bath spin have an exact analytical solution for the spin dynamics, while approximation-free results can be obtained numerically for sufficiently small systems with partially polarized nuclei\,\cite{JPCM.15.R1809}.

As we know, due to unavoidable interactions of the central spin with the surrounding nuclear spins, quantum coherence of the central spin will disappear finally. This decoherence process is the main obstacle to quantum computing and quantum information processing\,\cite{PRB.70.195340,PRB.76.045312,PRB.81.235324}. The environment destroys quantum interferences of the system, and the unwanted influences of the environment reduce the advantages of the quantum computing methods. Many studies were conducted to investigate the detrimental effect of decoherence without considering the interactions of the bath spins due to the complexity of the system. Another consideration is that, in quantum dots, the Heisenberg exchange resulting from the hyperfine interaction between the localized electron and the nuclei dominates on short-time scales up to 1 \rm{ms}\,\cite{Nature.435.925} before the interaction between the bath spins becomes effective.
It was found that the decoherence strongly depends on the homogeneity of the coupling between the central spin with its bath and the initial state of the bath spins\,\cite{PRL.88.186802,PRB.67.195329,JSM.2007.P06018,PRB.76.014304,PRB.82.161308}. The decoherence times become shorter for larger inhomogeneity in the couplings\,\cite{PRB.82.161308}. In the system with a fully polarized and untangled bath, a persistent oscillation of the central spin coherence, whose amplitude is inversely proportional to the total number of the bath spins, is discovered\,\cite{PRB.76.014304}, which means that the decoherence effect in quantum dots can be effectively suppressed by increasing the number of bath spins under the circumstance of initial spin polarization.

However, the bath spin interaction is ubiquitous in real materials. The bath spin coupling may induce a strong impact on the dynamical properties of some systems. {It was argued that energy exchange through mutual interactions of the bath modes avoids using the central spin as an intermediary and modifies the statistical properties of the energy levels and eigenstates of the bath, which can significantly affect the properties of the system\,\cite{JPA.36.12305}. However, it is still an open question to what extent the coupling of bath spins affects the decoherence and the possibility to tune the decoherence by external magnetic fields in an entangled environment.} Without considering the bath interactions, complete decoherence occurs in the central spin model with homogeneous couplings if the initially unentangled spin bath is with zero or small magnetization\,\cite{JSM.2007.P06018}. The central spin decays infinitely fast with increasing number of bath spins and to zero in the thermodynamical limit. However, for the environment initially in an eigenstate of the total bath spins, the central spin exhibits persistent monochromatic oscillation with a large amplitude\,\cite{JSM.2007.P06018}. An entangled bath state with zero or small magnetization may lead to persistent oscillations of the expectation value of the central spin in the homogeneous model, and very different long-time evolutions are induced by different degrees of entanglement\,\cite{PRB.76.014304}. It shows that the central spin coherence can be protected by entanglement in the bath in Refs.\,\cite{JPA.36.12305,PRA.71.052321,PRA.75.032333}. Moreover, it was found that the XY-type interaction among bath spins can induce a bath-size independent decay of the coherence\,\cite{PRA.75.032333}, which is also very different from that of the non-interacting cases. In the thermal dynamical limit, the analytical form of the Bloch vector for antiferromagnetic coupling within the spin bath was derived to investigate the short-time and long-time behavior of reduced dynamics of the central spin system\,\cite{PRB.76.174306}.

In this paper, we construct an extended solvable central spin model by introducing an isotropic {nearest-neighbour} spin-exchanging interaction {among} the bath spins into the standard central spin model\,\cite{JP.37.1087}, and we consider the finite external fields for the central spin and bath spins. By using the eigenvalues and Bethe eigenstates of the bath spins, we propose a method to solve this extended central spin model and obtain the exact solution of the system. By investigating the evolution of the central spin coherence and polarization, we show the rich dynamics of the system. The results obtained are helpful for understanding the decoherence problems in quantum dots. Meanwhile, we also find an interesting dephasing phenomenon.


The paper is organized as follows. In Sec.\,\ref{sec2}, we introduce the model Hamiltonian of the extended central spin system with the isotropic spin-exchanging interactions among the bath spins. In Sec.\,\ref{sec3.1}, we derive the exact solution of the system by using the Bethe ansatz and the theory of coupled angular momentum.
The reduced density matrix of the central spin is given in Sec.\,\ref{sec3.2}. In Sec.\,\ref{sec4}, we study the evolution of the central spin coherence, defined by the $l_1$ norm of coherence, from some specific initial states such as bath eigenstates and bath thermal states by using the measurement named $l_1$ norm. The dynamical properties of the central spin polarization are shown in Sec.\,\ref{sec5}. We give a brief summary in Sec.\,\ref{sec6}.

\section{The extended central spin model}
\label{sec2}

We consider a system in which one central spin is coupled with $N$ bath spins where the bath spins have the isotropic spin-exchange interactions. The model is schematically shown in Fig.1. For simplicity, we restrict ourselves to spins-$1/2$ objects. The Hamiltonian of this central spin model is
\begin{eqnarray}\label{eq2.1}
H=g\sum\limits_{j=1}^{N}\vec{S}\cdot\vec{\sigma}_j+ \Delta S^z+ \omega\sum\limits_{j=1}^{N}\sigma_j^z+H_x, \qquad H_{x}=\gamma\sum\limits_{j=1}^{N}\vec{\sigma}_j\cdot\vec{\sigma}_{j+1},
\end{eqnarray}
where $\vec{S}=\frac12(\sigma^x,\sigma^y,\sigma^z)$ denotes the central spin, $\vec{\sigma}_j=(\sigma_j^x,\sigma_j^y,\sigma_j^z)$ denotes
the $j$-th bath spin,
and $g$ is the isotropic coupling strength between the central spin and the surrounding bath spins.
The strengths of the external magnetic field measured by the central spin and the bath spins are $\Delta$ and $\omega$, respectively.
The last term $H_x$ describes the isotropic coupling between the bath spins, which is equivalent
to a periodic XXX spin-1/2 chain. The coupling of bath spins is ferromagnetic (antiferromagnetic) with $\gamma=1\ (\gamma=-1)$.
\begin{figure}[htbp]
\centering
\includegraphics[scale=0.4]{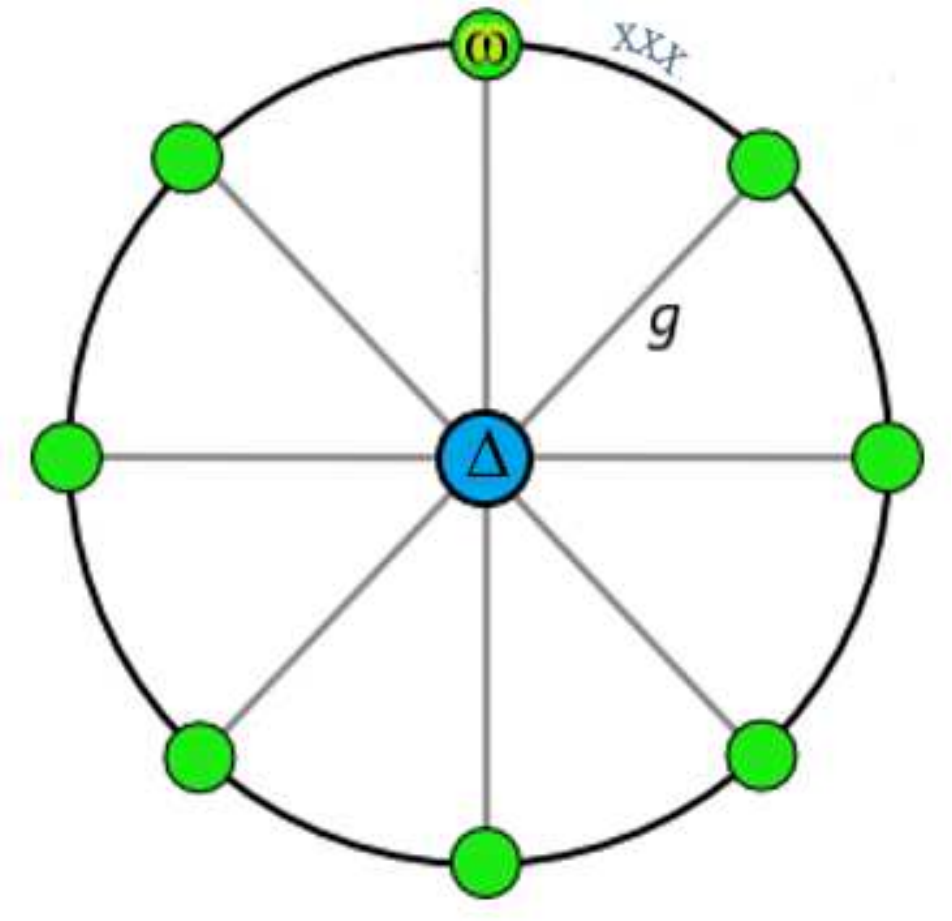}
  \caption{Schematic diagram of the extended central spin model with the isotropic exchanging interactions between the bath spins.}\label{fig1}
\end{figure}

Since the bath spins are identical and indistinguishable, we define the collective operators { $J^\alpha=\frac{1}{2}\sum\limits_{j=1}^{N}\sigma_j^\alpha$, with $\alpha=x,y, and z$}.
Then the spin-flipping operators are $J^\pm=J^x\pm iJ^y$. The spin-flipping operators acting on the central spin are defined by $S^\pm=S^x\pm iS^y$. Then the Hamiltonian\,(\ref{eq2.1}) can be rewritten as
\begin{eqnarray}\label{eq2.2}
H=g(J^+S^-+J^-S^++2J^zS^z)+\Delta S^z+\omega J^z+H_{x}.
\end{eqnarray}
The $su(2)$ symmetry of the Hamiltonian $H_x$ of the bath spins leads to $[H_x, J^\alpha]=0$. Meanwhile, $H_x$ and $S^\alpha$ are also commutative, i.e., $[H_x, S^\alpha]=0$. Therefore,
the Hamiltonian $H_{x}$ and $H$ commute with each other, i.e., $[H_x, H]=0$, which means
that they have common eigenstates.
We also note that if $\omega=\Delta=0$, the system (\ref{eq2.1}) has the $su(2)$ symmetry and the total spins (central spin plus bath spins) along each directions are all conserved, i.e., $[J^\alpha+S^\alpha,H]=0$. However, for the generic values of $\omega$ and $\Delta$, the $su(2)$ symmetry is broken and only the total spin along the $z$-direction is conserved, i.e., $[J^z+S^z,H]=0$.

\section{Exact solution and reduced density matrix}
\label{sec3}
\subsection{Exact solution}
\label{sec3.1}
In order to obtain the eigenvalues and the eigenstates of the Hamiltonnian (\ref{eq2.2}), we firstly seek the exact solution of the bath Hamiltonian $H_x$.
Using the language of the Bethe ansatz, $H_x$ is characterized by the $R$ matrix \,\cite{ODBA2015}
\begin{eqnarray}\label{eq3.1}
R_{0,j}(u)=u+P_{0,j}=u+\frac{1}{2}(1+\vec{\sigma}_j\cdot\vec{\sigma}_0),
\end{eqnarray}
where $u$ is the spectral parameter, $P_{0,j}$ is the permutation operator, $0$ means the auxiliary space, and $j$ means the quantum space. From the $R$ matrix, we can construct the monodromy matrix
\begin{eqnarray}\label{eq3.2}
T_0(u)=R_{0,N}(u)\cdots R_{0,1}(u)=\left(
                                     \begin{array}{cc}
                                       A(u) & B(u) \\
                                       C(u) & D(u) \\
                                     \end{array}
                                   \right).
\end{eqnarray}
It is remarked that the matrix elements $A(u)$, $B(u)$, $C(u)$, and $D(u)$ are the operators living in the Hilbert space of $H_x$. Taking the partial trace in the auxiliary space, we obtain the transfer matrix, i.e., $t(u)=tr_0T_0(u)$, which is the generating functions of all the conserved quantities of the
bath system. The Hamiltonian $H_x$ is the derivative of the logarithm of the transfer matrix $t(u)$:
{{\begin{eqnarray}\label{eq3.3}
H_x= 2\gamma\frac{d \ln t(u)}{d u}\big|_{u=0} + \gamma N.
\end{eqnarray}}}

The eigenstates of the Hamiltonian $H_x$ are constructed by enacting the spin-flipping operators $B(u)$ on the reference state $|\Omega\rangle$:
\begin{eqnarray}\label{eq3.4}
|\lambda_1,\cdots, \lambda_M\rangle=B(\lambda_1)\cdot\cdot\cdot B(\lambda_M)|\Omega\rangle, \qquad |\Omega\rangle=|\uparrow\rangle_1\otimes \cdots \otimes|\uparrow\rangle_N,
\end{eqnarray}
where \{$\lambda_j$\} are the Bethe roots, $M$ is the number of Bethe roots or the number of the flipped bath spins, and all the spins at the reference state $|\Omega\rangle$ are aligned as spin-up. The number of Bethe roots is set as $M\leq[\frac{N}{2}]$.
By using the algebraic Bethe ansatz method, the eigenvalues of $H_x$ are obtained as{{
\begin{eqnarray}\label{eq3.5}
E_x(\lambda_1,\cdots, \lambda_M)=-\sum\limits_{j=1}^{M}\frac{2\gamma}{\lambda_j^2+\frac{1}{4}}+\gamma N,
\end{eqnarray}}}
where the Bethe roots $\{\lambda_j\}$ satisfy the Bethe ansatz equations (BAEs)
\begin{eqnarray}\label{eq3.6}
\left(\frac{\lambda_j-\frac{i}{2}}{\lambda_j+\frac{i}{2}}\right)^N=\prod_{l\neq j}^M \frac{\lambda_j-\lambda_l-i}{\lambda_j-\lambda_l+i},\quad j=1,\ldots,M.
\end{eqnarray}

The Bethe-type eigenstates (\ref{eq3.4}) are the highest-weight states of the $J^+$ operator and are highly degenerate\,\cite{LNP.242.1985}. Because $[J^-, H_{x}]=0$, we can apply the $J^-$ operator to the Bethe states (\ref{eq3.4}) and obtain the complete basis of the Hilbert space of Hamiltonian $H_x$:
\begin{eqnarray}\label{eq3.7}
|\lambda_1,\cdots, \lambda_M; n\rangle \equiv (J^-)^{S_M-n}\frac{1}{Q_M}|\lambda_1,\cdots, \lambda_M\rangle, \qquad n=S_M, \cdots,  -S_M,\label{state}
\end{eqnarray}
where $S_M=N/2-M$ is the magnetization of the Bethe state $|\lambda_1,\cdots, \lambda_M\rangle$ and $Q_M=\sqrt{\langle \lambda_1,\cdots, \lambda_M |\lambda_1,\cdots, \lambda_M\rangle}$ is the corresponding normalization factor. The basis set (\ref{state}) satisfies the following
\begin{eqnarray}
&&J^-|\lambda_1,\cdots, \lambda_M;-S_M\rangle=0, \nonumber  \\
&&J^-|\lambda_1,\cdots, \lambda_M; n\rangle=\sqrt{(S_M+n)(S_M+1-n)}|\lambda_1,\cdots, \lambda_M; n-1\rangle,\nonumber  \\
&&J^+|\lambda_1,\cdots, \lambda_M; n\rangle=\sqrt{(S_M+1+n)(S_M-n)}|\lambda_1,\cdots, \lambda_M; n+1\rangle, \nonumber \\
&&J^z|\lambda_1,\cdots, \lambda_M; n\rangle=n|\lambda_1,\cdots, \lambda_M; n\rangle.\label{eq3.8}
\end{eqnarray}
The basis with the same $M$ but different $n$ is the degenerated eigenstates of the Hamitonian $H_x$, and the corresponding energy is $E_x(\lambda_1,\cdots, \lambda_M)$.
They are uniquely determined by the complete set of physical quantities $H_x$ and $J^-$.
The completeness of them has been proven by Tarasov in Ref.\,\cite{RevMathPhys.30.1840018}.

Now, we are ready to solve the extended central spin system (\ref{eq2.2}). First, we construct the complete basis of the Hilbert space of Hamiltonian (\ref{eq2.2}) as
\begin{eqnarray}\label{eq3.9}
&&\{|\lambda_1,\cdots, \lambda_M; n\rangle\otimes|\uparrow\rangle,\quad |\lambda_1,\cdots, \lambda_M; n\rangle\otimes|\downarrow\rangle\},  \nonumber \\
&&  n=S_M, \cdots, -S_M, \quad M=0,1, \cdots, [\frac{N}2],
\end{eqnarray}
where $|\uparrow\rangle$ and $|\downarrow\rangle$ are the basis of the central spin.
By acting Hamiltonian (\ref{eq2.2}) on the above basis (\ref{eq3.11}), we get
\begin{eqnarray}
&&H|\lambda_1, \cdots, \lambda_M;S_M\rangle\otimes|\uparrow\rangle= (E_x+\omega S_M+\frac{\Delta}{2}+gS_M)|\lambda_1,\cdots, \lambda_M;S_M\rangle\otimes|\uparrow\rangle,\label{eq3.10} \\
&&H|\lambda_1, \cdots, \lambda_M;-S_M\rangle\otimes|\downarrow\rangle= \hspace{-2.5pt}(E_{x}\hspace{-2.5pt}-\hspace{-2.5pt}\omega S_M-\frac{\Delta}{2}+gS_M)|\lambda_1,\cdots, \lambda_M;-S_M\rangle\otimes|\downarrow\rangle,\label{eq3.11}\\
&&H|\lambda_1, \cdots, \lambda_M; m\rangle\otimes|\uparrow\rangle=(E_{x}+m\omega +\frac{\Delta}{2}+m g )|\lambda_1,\cdots, \lambda_M; m\rangle\otimes|\uparrow\rangle
\nonumber \\
&&\qquad  +g\sqrt{(S_M+1+m)(S_M-m)}|\lambda_1,\cdots, \lambda_M; m+1\rangle\otimes|\downarrow\rangle,\quad m\neq S_M, \label{eq3.12} \\
&&H|\lambda_1, \cdots, \lambda_M; m+1\rangle\otimes|\downarrow\rangle=(E_{x}+(\omega-g) (m+1)-\frac{\Delta}{2})|\lambda_1, \cdots, \lambda_M; m+1\rangle\otimes|\downarrow\rangle\nonumber \\
&&\qquad +g\sqrt{(S_M+1+m)(S_M-m)}|\lambda_1, \cdots, \lambda_M; m\rangle\otimes|\uparrow\rangle, \quad m\neq S_M, \label{eq3.13}
\end{eqnarray}
where $E_x=E_x(\lambda_1,\cdots, \lambda_M)$.

From Eqs.(\ref{eq3.10}) and (\ref{eq3.11}), we obtain two eigenstates of the central spin system directly as
\begin{eqnarray}\label{eq3.14}
&&|\lambda_1,\cdots, \lambda_M;S_M\rangle\otimes|\uparrow\rangle, \nonumber \\
&&  |\lambda_1,\cdots, \lambda_M;-S_M\rangle\otimes|\downarrow\rangle,
\end{eqnarray}
and we obtain the corresponding eigenvalues in the form
\begin{eqnarray}\label{eq3.15}
E_{S_M}^{\pm}=E_{x}(\lambda_1,\cdots, \lambda_M)+gS_M\pm(\omega S_M+\frac{\Delta}{2}).
\end{eqnarray}

From Eqs.(\ref{eq3.12}) and (\ref{eq3.13}), we find that the basis
\begin{eqnarray}\label{eq3.16}
&& |\lambda_1, \cdots, \lambda_M; m\rangle\otimes|\uparrow\rangle, \nonumber \\
&& |\lambda_1, \cdots, \lambda_M; m+1\rangle\otimes|\downarrow\rangle,
\end{eqnarray}
constructs a two-dimensional invariant subspace. Therefore,
the Hilbert space of the system can be divided as the direct product of these kinds of invariant subspaces, and
the Hamiltonian (\ref{eq2.2}) is the direct sum of determined 2$\times$2 matrix blocks. This means that we can diagonalize the Hamiltonian in every subblock. In the invariant subspace (\ref{eq3.14}), the matrix form of the Hamiltonian is
\begin{eqnarray}\label{eq3.17}
&&\left(
    \begin{array}{cc}
      E_{x}(\lambda_1,\cdots, \lambda_M)+m(g+\omega)+\frac{\Delta}{2} & g\sqrt{(S_M+1+m)(S_M-m)} \\[6pt]
      g\sqrt{(S_M+1+m)(S_M-m)} & E_{x}(\lambda_1,\cdots, \lambda_M)+(m+1)(\omega-g)-\frac{\Delta}{2}
    \end{array}
  \right), \nonumber
\end{eqnarray}
where $m\neq S_M$. By solving the resulting secular equation, we obtain the eigenvalues of the Hamiltonian (\ref{eq2.2}) as
\begin{eqnarray}\label{eq3.18}
E_m^{\pm}&=&E_{x}(\lambda_1,\cdots, \lambda_M)+m\omega\notag +\frac12\left\{\omega-g\pm\left[
[w-\Delta-g(2m+1)]^2\right.\right. \nonumber \\ &&\left.\left. +4g^2(S_M+1+m)(S_M-m)\right]^{1/2}\right\},
\end{eqnarray}
where $m\in[-S_M, S_M-1]$. The corresponding eigenstates are
\begin{eqnarray}\label{eq3.19}
&&g_m|\lambda_1,\cdots, \lambda_M; m\rangle \otimes| \uparrow\rangle
+ ( \varepsilon_m \pm \zeta_m)|\lambda_1,\cdots, \lambda_M;
m+1\rangle \otimes |\downarrow\rangle,
\end{eqnarray}
where the parameters $\varepsilon_m$, $g_m$ and $\zeta_m$ are given by
\begin{eqnarray}\label{eq3.20}
&&\varepsilon_m=[\omega-\Delta-g(2m+1)]/2, \nonumber \\
&&g_m=g\sqrt{(S_M+1+m)(S_M-m)},\nonumber \\
&&\zeta_m=\sqrt{\varepsilon_m^2+g_m^2}.\label{eq3.21}
\end{eqnarray}
respectively.

\subsection{Reduced density matrix}
\label{sec3.2} 

In order to study the dynamical evolution of the system, we calculate the time-dependent reduced density matrix of the central spin by using
the eigenvalues and eigenstates obtain in Sec.\,\ref{sec3.1}. The reduced density matrix is defined as
\begin{eqnarray}\label{eq4.1}
\rho_{c}(t)=tr_B\{U(t)\rho_{tot}(0)U(t)^{\dag}\},
\end{eqnarray}
where $tr_B$ means taking the partial trace of the bath spins, $U(t)=\exp(-iHt)$ is the time evolution operator, and $\rho_{tot}(0)$ is the initial density matrix of the system described by the Hamiltonian (\ref{eq2.2}). Hereafter, we consider an initial state
\begin{eqnarray}\label{eq4.2}
\rho_{tot}(0)=\rho_{c}(0)\otimes\rho_{B}(0),
\end{eqnarray}
where $\rho_{c}(0)$ and $\rho_{B}(0)$ are the initial density matrices of the central spin and the bath spins, respectively. Obviously, at the initial state, the
central spin and the bath spins do not have correlations.
We note that $\rho_{tot}(0)$ ensures the complete positivity of the reduced dynamics \,\,\cite{JMP.17.821,CMP.48.119}.

At the initial time $t=0$, we suppose that the state of the central spin is a general superposition
state, $\rho_c(0)=(\alpha|\uparrow\rangle+\beta|\downarrow\rangle)(\langle\uparrow|\alpha^*+\langle\downarrow|\beta^*)$, where $\alpha$ and $\beta$ are
the coefficients satisfying the normalization $|\alpha|^2+|\beta|^2=1$.
Then, we should determine the initial states of the bath. We consider two fundamental cases. One is the bath initially occupying its own eigenstate,
$\rho_{B}(0)=|\lambda_1, \cdots, \lambda_M; n\rangle\langle\lambda_1, \cdots, \lambda_M; n|$,
and the other is the bath initially in its own thermal state,
$\rho_B(0)=\frac{1}{Z}e^{-H_{x}/(k_BT)}$, at a finite temperature $T$, where $k_B$ is the Boltzman constant and $Z$ is the partition function,
\begin{eqnarray}\label{eq4.3}
Z=\sum_{M=0}^{[\frac{N}{2}]}\sum_{\{\lambda_1, \cdots, \lambda_M\}}\sum_{n=-S_M}^{S_M}e^{-\frac{E_x(\lambda_1,\cdots, \lambda_M)}{k_BT}}.
\end{eqnarray}

Note that the initial state (\ref{eq4.2}) is not the eigenstate of the central spin system (\ref{eq2.2}),
and we need to expand it by the eigenstates (\ref{eq3.16}) and (\ref{eq3.19}).
From the definition (\ref{eq4.1}) and using the initial condition (\ref{eq4.2}) as well as the corresponding eigenvalues (\ref{eq3.15}) and (\ref{eq3.18}), the reduced density matrix of the central spin at time $t$ can be obtained analytically. {In previous study\,\cite{PRB.76.174306}, the analytical form of the Bloch vector was derived for antiferromagnetic interactions within the bath in the limit of an infinite number of environmental spins. In inhomogeneous central-spin models without bath spin coupling, analytical results were derived in the limiting case of a fully polarized bath\,\cite{PRB.67.195329,PRB.76.014304}.}

If the initial bath state is the highest weight state $|\lambda_1, \cdots, \lambda_M; S_M\rangle$, with $S_M=0$, the elements of the reduced density matrix of the central spin at time $t$ are
\begin{eqnarray}\label{eq4.4}
\rho_{c,M,0}^{11}(t)=1-\rho_{c,M,0}^{22}(t)=|\alpha|^2, \quad
\rho_{c,M,0}^{12}(t)=\rho_{c,M,0}^{21}(t)^*=\alpha\beta^*e^{-i\Delta t}\equiv \alpha\beta^*r_{M,0}(t),
\end{eqnarray}
while with $S_M\neq0$, the elements of the reduced density matrix read
\begin{eqnarray}\label{eq4.5}
\rho_{c,M,S_M}^{11}(t)&=&1-\rho_{c,M,S_M}^{22}(t)=|\alpha|^2+\frac{|\beta|^2}{2\zeta_{S_M-1}^2}g_{S_M-1}^2[1-\cos(\zeta_{S_M-1}t)],\nonumber \\
\rho_{c,M,S_M}^{12}(t)&=&\rho_{c,M,S_M}^{21}(t)^*=\alpha\beta^*\frac{e^{-i\omega t}}{2\zeta_{S_M-1}}[(\varepsilon_{S_M-1}+\zeta_{S_M-1})e^{i(\varepsilon_{S_M}+\zeta_{S_M-1})t}\nonumber
\\[2pt]
&&-(\varepsilon_{S_M-1}-\zeta_{S_M-1})e^{-i(-\varepsilon_{S_M}+\zeta_{S_M-1})t}]\equiv \alpha\beta^*r_{M,S_M}(t).
\end{eqnarray}

If the initial bath state is the lowest weight state $|\lambda_1, \cdots, \lambda_M; -S_M\rangle$, the elements of the reduced density matrix are
\begin{eqnarray}\label{eq4.6}
\rho_{c,M,-S_M}^{11}(t)&=&1-\rho_{c,M,-S_M}^{22}(t)=\frac{|\alpha|^2}{2\zeta_{-S_M}^2}[2\varepsilon_{-S_M}^2+g_{-S_M}^2(1+\cos(\zeta_{-S_M}t))]+|\beta|^2,\nonumber \\
\rho_{c,M,-S_M}^{12}(t)&=&\rho_{c,M,-S_M}^{21}(t)^*=\alpha\beta^*\frac{e^{-i\omega t}}{2\zeta_{-S_M}}[(-\varepsilon_{-S_M}+\zeta_{-S_M})e^{-i(-\varepsilon_{-S_M+1}+\zeta_{-S_M})t}\nonumber
\\[2pt]
&&+(\varepsilon_{-S_M}+\zeta_{-S_M})e^{i(\varepsilon_{-S_M+1}+\zeta_{-S_M})t}]\equiv \alpha\beta^*r_{M,-S_M}(t).
\end{eqnarray}

The elements of the reduced density matrix of the central spin with the bath occupying other eigenstates $|\lambda_1, \cdots, \lambda_M; m^{\prime}\rangle$ are
\begin{eqnarray}\label{eq4.7}
\rho_{c,M,m^{\prime}}^{11}(t)&=&1-\rho_{c,M,m^{\prime}}^{22}(t)=\frac{|\alpha|^2}{2\zeta_{m^{\prime}}^2}[2\varepsilon_{m^{\prime}}^2+g_{m^{\prime}}^2(1+\cos(\zeta_{m^{\prime}}t))]+\frac{|\beta|^2}{2\zeta_{{m^{\prime}}-1}^2}g_{{m^{\prime}}-1}^2[1-\cos(\zeta_{{m^{\prime}}-1}t)],\nonumber \\
\rho_{c,M,{m^{\prime}}}^{12}(t)&=&\rho_{c,M,{m^{\prime}}}^{21}(t)^*\nonumber \\
&=&\alpha\beta^* \frac{e^{-i\omega t}}{4\zeta_{m^{\prime}}\zeta_{{m^{\prime}}-1}}[(-\varepsilon_{m^{\prime}}+\zeta_{m^{\prime}})(\varepsilon_{{m^{\prime}}-1}+\zeta_{{m^{\prime}}-1})e^{-i(\zeta_{m^{\prime}}-\zeta_{{m^{\prime}}-1})t}-(\varepsilon_{m^{\prime}}+\zeta_{m^{\prime}})\nonumber
\\&&\times (\varepsilon_{{m^{\prime}}-1}-\zeta_{{m^{\prime}}-1})e^{-i(-\zeta_{m^{\prime}}+\zeta_{{m^{\prime}}-1})t}-(-\varepsilon_{m^{\prime}}+\zeta_{m^{\prime}})(\varepsilon_{{m^{\prime}}-1}-\zeta_{{m^{\prime}}-1})e^{-i(\zeta_{m^{\prime}}+\zeta_{{m^{\prime}}-1})t}\nonumber
\\&&+(\varepsilon_{m^{\prime}}+\zeta_{m^{\prime}})(\varepsilon_{{m^{\prime}}-1}+\zeta_{{m^{\prime}}-1})e^{i(\zeta_{m^{\prime}}+\zeta_{{m^{\prime}}-1})t}]
\equiv \alpha\beta^*r_{M,{m^{\prime}}}(t),
\end{eqnarray}
with ${m^{\prime}}\in[-S_M+1,S_M-1]$.
For the initial bath spins at the thermal state,
the elements of the reduced density matrix are
\begin{eqnarray}\label{eq4.8}
\rho_c^{kl}(t)=\frac{1}{Z} \sum_{M=0}^{[\frac{N}{2}]}\sum_{\{\lambda_1, \cdots, \lambda_M\}}\sum_{n=-S_M}^{S_M} e^{-\frac{E_x(\lambda_1,\cdots, \lambda_M)}{K_BT}}\rho^{kl}_{c,M,n}(t),
\end{eqnarray}
where $k, l=\{1,2\}$ and $\rho^{kl}_{c,M,n}(t)$ are given by Eqs.(\ref{eq4.4})-(\ref{eq4.7}).

\section{Evolution of the central spin coherence}
\label{sec4} 

Provided the central spin is a qubit, the evolution of the quantum coherence of the central spin is characterized by the so-called $l_1$ norm of the coherence $C_{l_1}$ determined by the absolute value of the off-diagonal element
$\rho_c^{12}(t)$ of the reduced density matrix\,\cite{PRL.113.140401},
\begin{eqnarray}\label{eq5.1}
C_{l_1}(t)=2|\rho_c^{12}(t)|=2|\alpha \beta^*|| r_{M, n}|,
\end{eqnarray}
which makes it possible to study the evolution of the coherence with different model parameters, showing some interesting results.

As the bath spins supply the environment of the central spin, the evolution of the central spin coherence shows varying properties with different initial bath states. Thus we consider the initial bath eigenstates (including the ground state) and the initial bath thermal states separately. Meanwhile, we require that the combination coefficients $\alpha$ and $\beta$ of the central spin cannot be zero, i.e., $\alpha\beta\neq0$.
Otherwise the central spin only occupies the spin-up state or the spin-down state which does not evolve.
From Eq.\,(\ref{eq5.1}), we also know that the quantity $|\alpha\beta|$ only contributes a time-independent constant factor and does not contribute to the evolution process. So we only consider the coherence factor $|r_{M, n}|$ in the following.

\subsection{Bath initially in its own eigenstates}
\label{sec4.1}

Under the condition that all the bath spins occupy their eigenstates, if the bath eigenstates are the highest-weight (lowest-weight) states $|\lambda_1,\cdots, \lambda_M; S_M\rangle$ with $S_M=0$, we know $r_{M,0}(t)=e^{-i\Delta t}$, which is a phase factor from Eq.\,(\ref{eq4.4}). Thus the coherence factor is
\begin{eqnarray}
|r_{M,0}(t)|=1,
\end{eqnarray}
which means that the coherence of the central spin does not evolve. If the bath eigenstates are the
highest-weight state $|\lambda_1,\cdots, \lambda_M; S_M\rangle$ or the lowest weight state $|\lambda_1,\cdots, \lambda_M; -S_M\rangle$
with $S_M\neq 0$, the coherence factors are
\begin{eqnarray}\label{eq5.2}
|r_{M, S_M}(t)|=f_{M, S_M-1}(t), \quad |r_{M, -S_M}(t)|=f_{M,-S_M}(t).
\end{eqnarray}
If the bath eigenstates $|\lambda_1,\cdots, \lambda_M; m^{\prime}\rangle$ with ${m^{\prime}}\neq\pm S_M$ are between the highest-weight and the lowest-weight states,
the coherence factor is
\begin{eqnarray}\label{eq5.3}
|r_{M, {m^{\prime}}}(t)|=f_{M, {m^{\prime}}}(t)f_{M, {m^{\prime}}-1}(t),\quad  {m^{\prime}}\in[-S_M+1,S_M-1].
\end{eqnarray}
The function $f_{M, n}(t)$ is given by
\begin{eqnarray}\label{eq5.4}
f_{M,n}(t)=\left\{1-\frac{g_n^2}{\zeta_n^2}\sin^2(\zeta_nt)\right\}^{\frac12}, \quad n=S_M, \cdot\cdot\cdot, -S_M.
\end{eqnarray}

\begin{figure}[htbp]
  \centering
  \subfigure{
    \label{fig2:subfig:a} 
    \includegraphics[width=3.3in, height=2.5in]{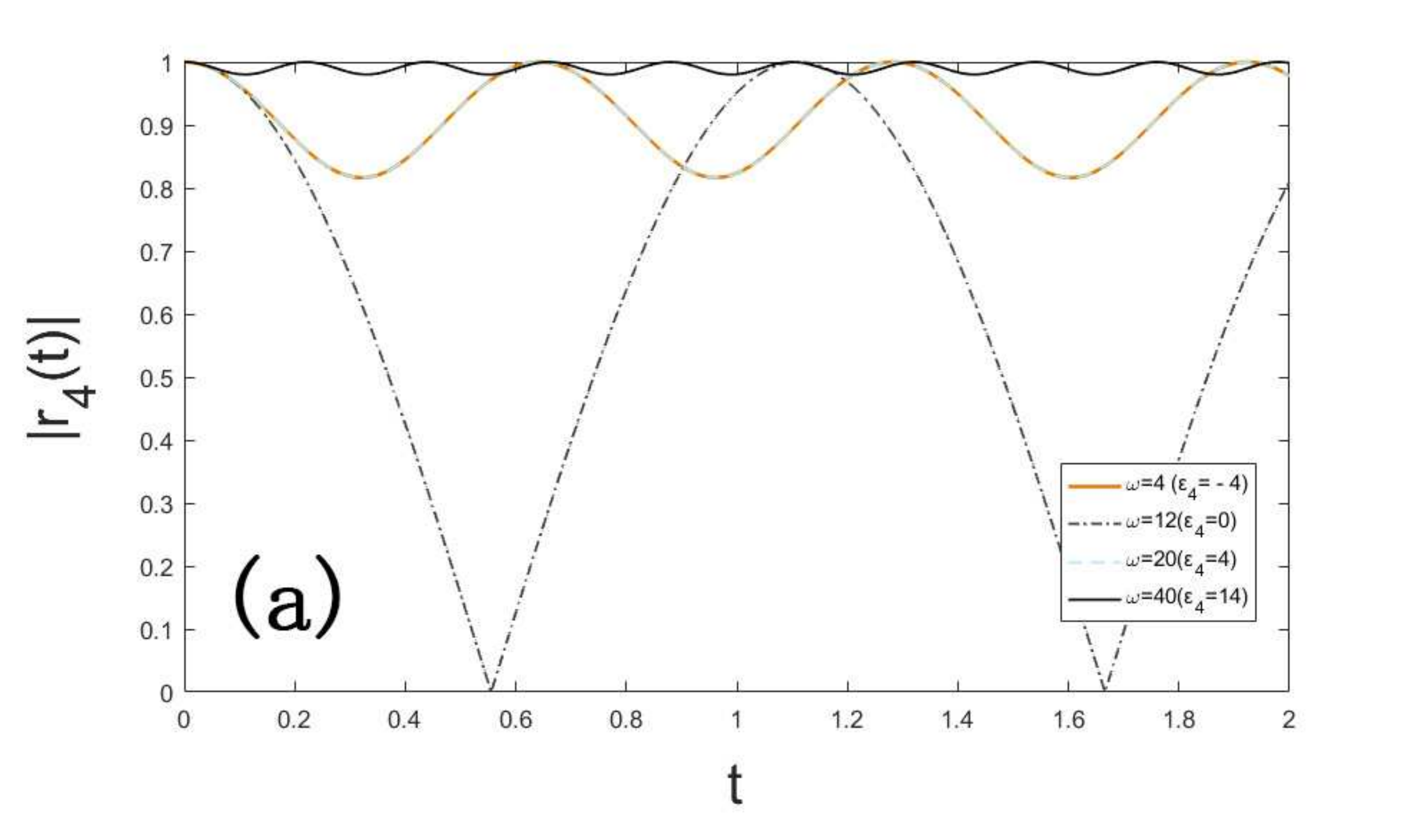}
  }
  \hfill
  \subfigure{
    \label{fig2:subfig:b} 
    \includegraphics[width=3.3in, height=2.5in]{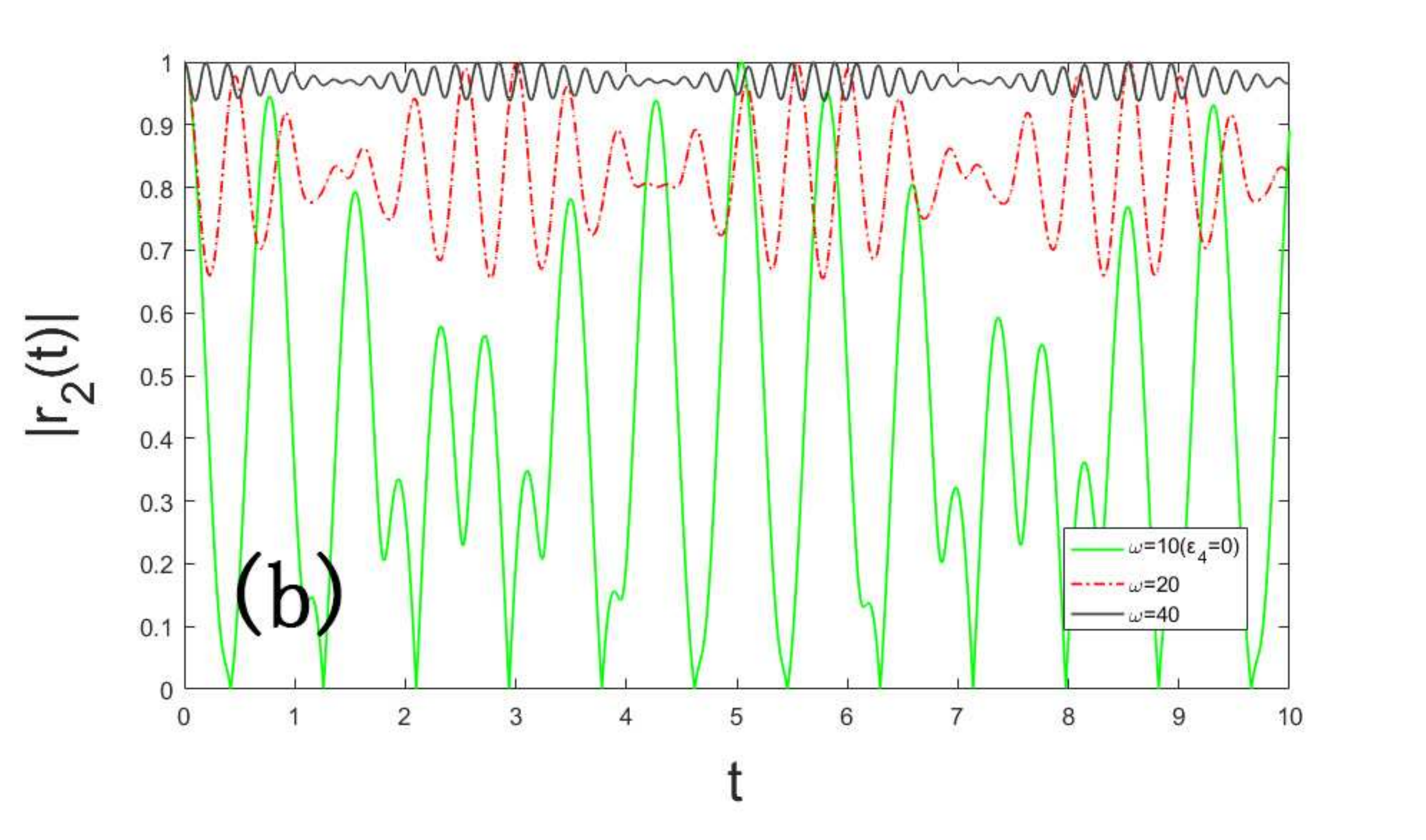}
  }
  \caption{Real-time evolutions of the coherence factor of the central spin when the bath is in a highest-weight state with $S_M=n=4$ (a) and a state between the highest and the lowest weight with $S_M=4$ and $n=2$ (b) for different $\omega$. In both plots, we set $g=1$, $\Delta=5$, and $N=8$.}
\end{figure}

From the parametrization (\ref{eq3.21}) and the definition (\ref{eq5.3}), we see that the coherence factor $|r_{M, m}(t)|$ depends on the quantity $(\omega-\Delta)$ and $g$, while it is independent of $E_x$. Hence we keep $\Delta=5$ a constant and tune the value of $\omega$. The real-time evolutions of the central spin coherence induced by a highest-weight state with $S_M=n=4$ and by a non-highest-weight state with
$S_M=4$ and $n=2$ for fixed $g$ are shown in Figs.\,\ref{fig2:subfig:a} and\,\ref{fig2:subfig:b}, respectively.

In Fig.\,\ref{fig2:subfig:a}, we show the oscillation behavior (i.e., $|\cos(gt)|$) of the central spin coherence when $\varepsilon_{S_M-1}=0$ corresponding to the highest-weight state as described by Eq.\,(\ref{eq5.2}). For the lowest-weight state, the coherence has the same property as for the highest one. However, as shown in Fig.\,\ref{fig2:subfig:b}, this kind of oscillation behavior does not exist for other states due to the fact that $\varepsilon_{m^{\prime}}=0$ and $\varepsilon_{{m^{\prime}}-1}=0$ cannot be $0$ at the same time. Then the coherence factors for these states are the product of two parts according to Eq.\,(\ref{eq5.3}).

As we know, the initial state is not an eigenstate of the Hamiltonian (\ref{eq2.2}).
During the evolution, if the bath spins occupy the highest-weight (lowest-weight) states, only the states with $S_M$ and $S_M-1$ ($-S_M$ and $-S_M+1$) contribute to the central spin coherence according to Eqs.\,(\ref{eq4.5}) and \,(\ref{eq4.6}). All the other states except for those with the highest weight and the lowest weight evolves in three state subspaces with ${m^{\prime}}$ and ${m^{\prime}}\pm 1$ according to Eq.\,(\ref{eq4.7}).

\begin{figure}[t]
  \centering
  \subfigure{
    \label{fig3:subfig:a} 
    \includegraphics[width=3.3in, height=2.5in]{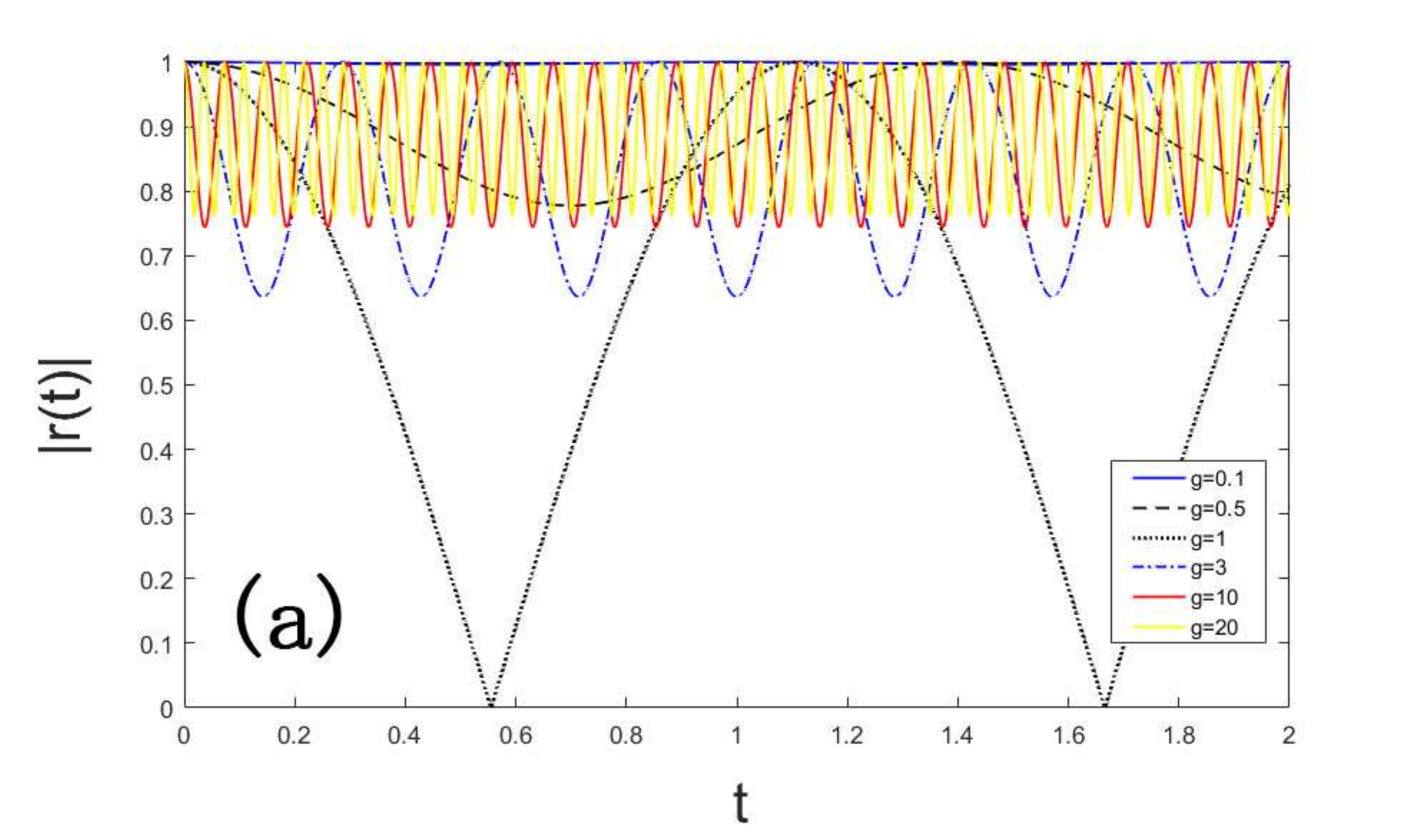}
  }
  \hfill
  \subfigure{
    \label{fig3:subfig:b} 
    \includegraphics[width=3.3in, height=2.5in]{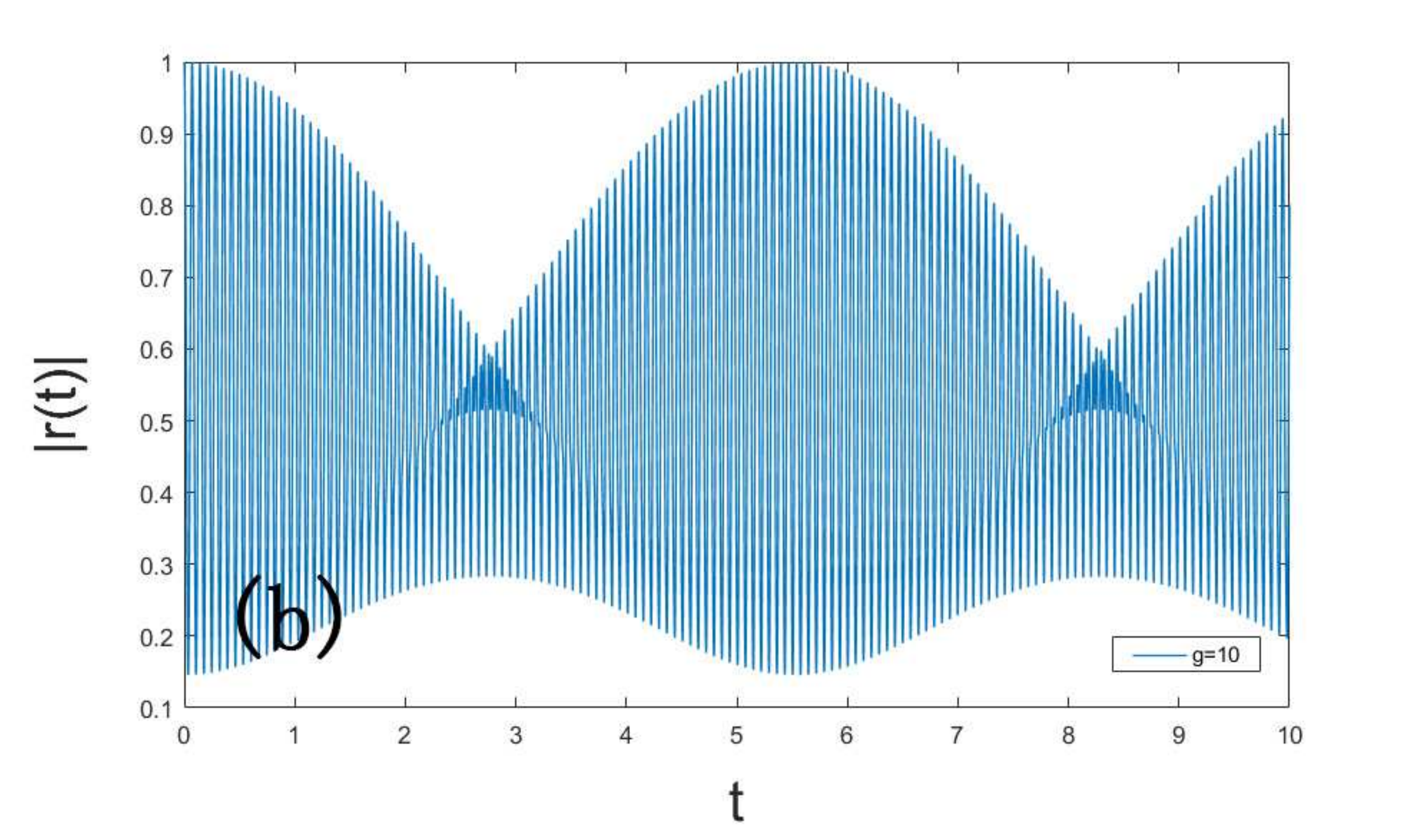}
  }
   \caption{Real-time evolutions of the coherence factor of the central spin when the bath is in a highest-weight state with $S_M=n=4$ for different $g$ (a), and in a state between the highest and the lowest weight with $S_M=4$ and $n=2$ for $g=10$ (b). In both plots, we set $\omega=12$, $\Delta=5$, and $N=8$.}\label{fig3}
\end{figure}

As shown in Figs.\,\ref{fig2:subfig:a} and \ref{fig2:subfig:b}, the lower bound of the coherence factor $|r_{M, n}(t)|$ and the oscillation frequency increase with increasing $|\omega-\omega_{c}|$. There is a critical external magnetic field, $\omega_{c}$, applied to the bath spins determined by the constraint $\varepsilon_{S_M-1}=0$, which makes the lower bound of the coherence factor $|r_{M, n}(t)|$ reach $0$. When $|\omega-\omega_{c}|$ increases, the factor $g_n/\zeta_n$ in Eq.\,(\ref{eq5.4}) turns out to be smaller than $1$, which raises up the lowest bound of the coherence factor. If $|\omega-\omega_{c}|$ is big enough, the coefficient $g_n/\zeta_n$ will be close to 0 and the coherence factor $|r_{M, n}(t)|$ will approach 1, which means the central spin coherence is kept very well.
Now, we study the evolution of the central spin coherence
for fixed $\omega$. The evolutions of the coherence factor $|r_{M, n}(t)|$ induced by the highest-weight (lowest-weight) state and other states of the bath spins are shown in Figs.\,\ref{fig3:subfig:a} and \ref{fig3:subfig:b}, respectively.
We see that the lower bound of the coherence factor $|r_{M, n}(t)|$ moves up
with increasing $g$ from the value $g_{c}$, where
$g_{c}$ is the critical coupling between the central spin and the bath spins determined by the constraint $\varepsilon_{S_M-1}=0$. Here, with increasing $g$,
the increasing rate of the lower bound of $|r_{M, n}(t)|$ turns slowly, while the oscillation of $|r_{M, n}(t)|$ becomes rapidly.
These phenomena can be explained analytically based on the knowledge given in Sec.\,\ref{sec3.1}.
With increasing $g$, both $\varepsilon_n$ and $g_n$ grow. Thus the oscillation frequency $\zeta_n$ increases while the oscillation amplitude decreases. Meanwhile, the increasing of $\varepsilon_n$ will enhance the coherence while the increasing of $g_n$ will reduce the coherence.
Due to the fact that $g_n$ will decrease with decreasing $g$ from the critical value of $g_{c}$, we can draw the conclusion that the weak coupling between the central spin and the bath spins can enhance the central spin coherence.

The physical picture is as follows. Although the interactions between the central spin and the bath spins are isotropic, from the derivation in Sec.\,\ref{sec3.1}, we see that the contribution of the couplings between the central spin and the bath spins in the $z$ direction enters the term $\varepsilon_n$, while the contributions of the couplings in the $x$ and $y$ directions enter the term $g_n$. From Eq.\,(\ref{eq3.21}), the couplings in the $z$ direction can be regarded as a total effective magnetic field applied on the central spin. This total effective magnetic field is generated by the polarization of bath spins as well as the external magnetic field applied on the central spin, which tunes the energy gap of the central spin. It is hard to flip the central spin from the spin-up state to the spin-down state if the total effective magnetic field is very large. However, the couplings in the $x$ and $y$ directions will reduce the coherence of the central spin. This is because more possible states are involved during the evolution, which can be seen from Eqs.\,(\ref{eq3.12}) and (\ref{eq3.13}).

We also find that for the highest-weight states with different $S_M$, the lower bound of the coherence factor $|r_{M, S_M}(t)|$ is smaller if $S_M$ is larger. This is because, for a large $S_M$, $g_n$ is large and more possible states contribute to the coherence of the central spin. Therefore, we conclude that the coherence of the central spin can be enhanced by increasing $|\omega-\omega_{c}|$ or decreasing $|g-g_c|$ in the system (\ref{eq2.1}).

If all the bath spins occupy their eigenstates, although there are interactions between the bath spins, the coupling constant $\gamma$, the energy $E_{x}$ of the bath spins, and some details of Bethe states are still not included in the coherence factor $|r_{M, n}(t)|$. This is because the operation of taking trace erases some information about the bath spins. If we consider other physical quantities or the evolution of the coherence at the thermal state, the corresponding results will be different as discussed in the following.

\subsection{Bath initially in its own thermal state}
\label{sec4.2}
In order to study the effects induced by the interaction of the bath spins, we consider that the evolution of the central spin coherence begins in a thermal environment at a finite temperature $T$. All the eigenstates $|\lambda_1,\cdots, \lambda_M; S_M\rangle$ with degeneracy $S_M$ of bath spins should be ergodic during the evolution. Then the central spin coherence is the summation of the production of the coherence of each eigenstate, $r_{M, n}(t)$, and its weight, $e^{-E_x/(k_BT)}$, for all the eigenstates of the bath spins in the form of
\begin{eqnarray}\label{eq5.5}
|r(t)|=\left|\frac{1}{Z}\sum_{M=0}^{[\frac{N}{2}]}\sum_{\{\lambda_1, \cdots, \lambda_M\}}\sum_{n=-S_M}^{S_M}e^{-\frac{E_x(\lambda_1,\cdots, \lambda_M)}{k_BT}}r_{M, n}(t)\right|,
\end{eqnarray}
where the energy $E_x$ of the bath spins is determined by the Bethe roots $\{\lambda_1,\cdots, \lambda_M\}$ obtained from Eq.\,(\ref{eq3.4}) and by solving the BAEs\,(\ref{eq3.5}).

\begin{figure}[htbp]
  \centering
  \subfigure{
    \label{fig4:subfig:a} 
    \includegraphics[width=3.4in, height=2.6in]{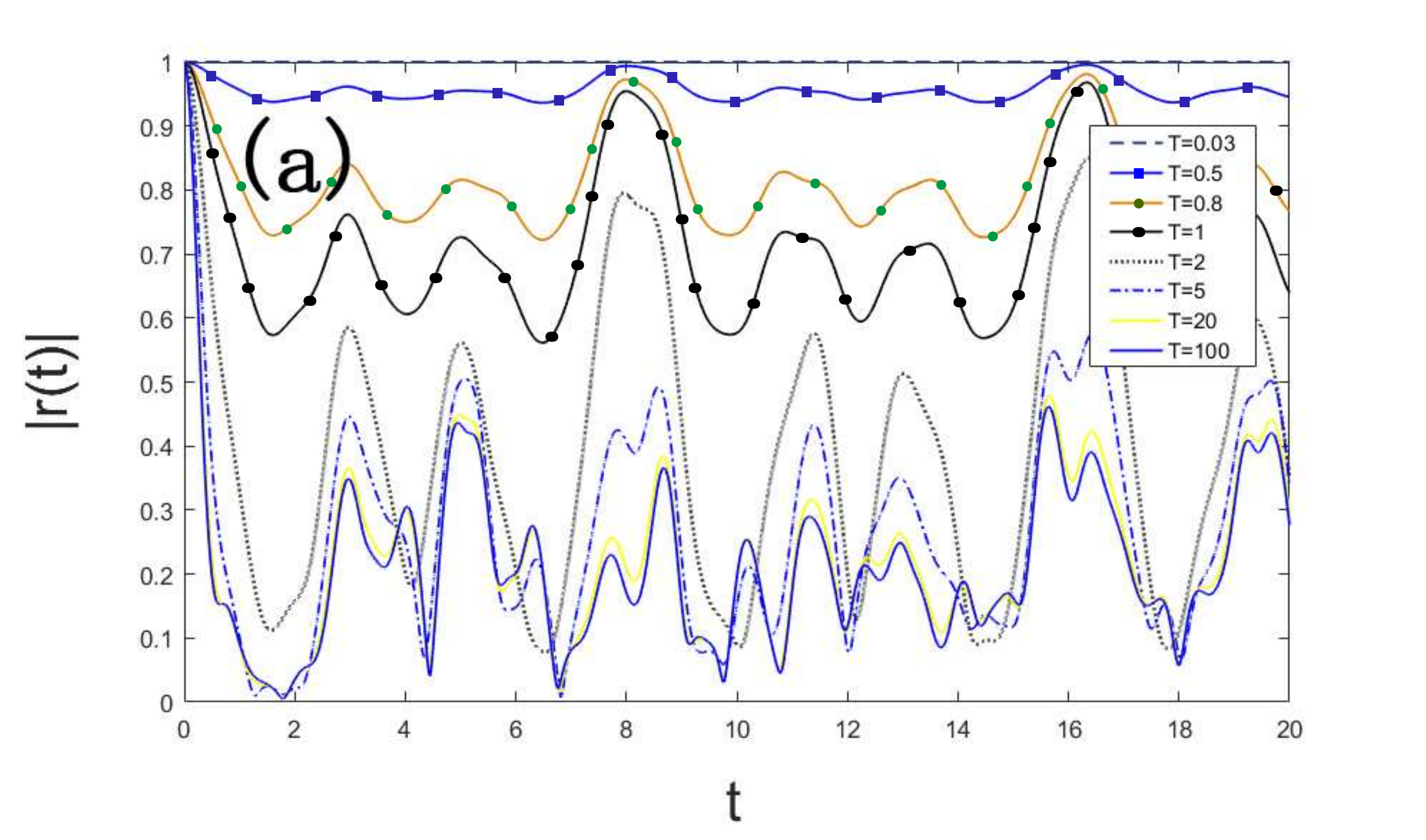}
  }
  \hfill
  \subfigure{
    \label{fig4:subfig:b} 
    \includegraphics[width=3.4in, height=2.6in]{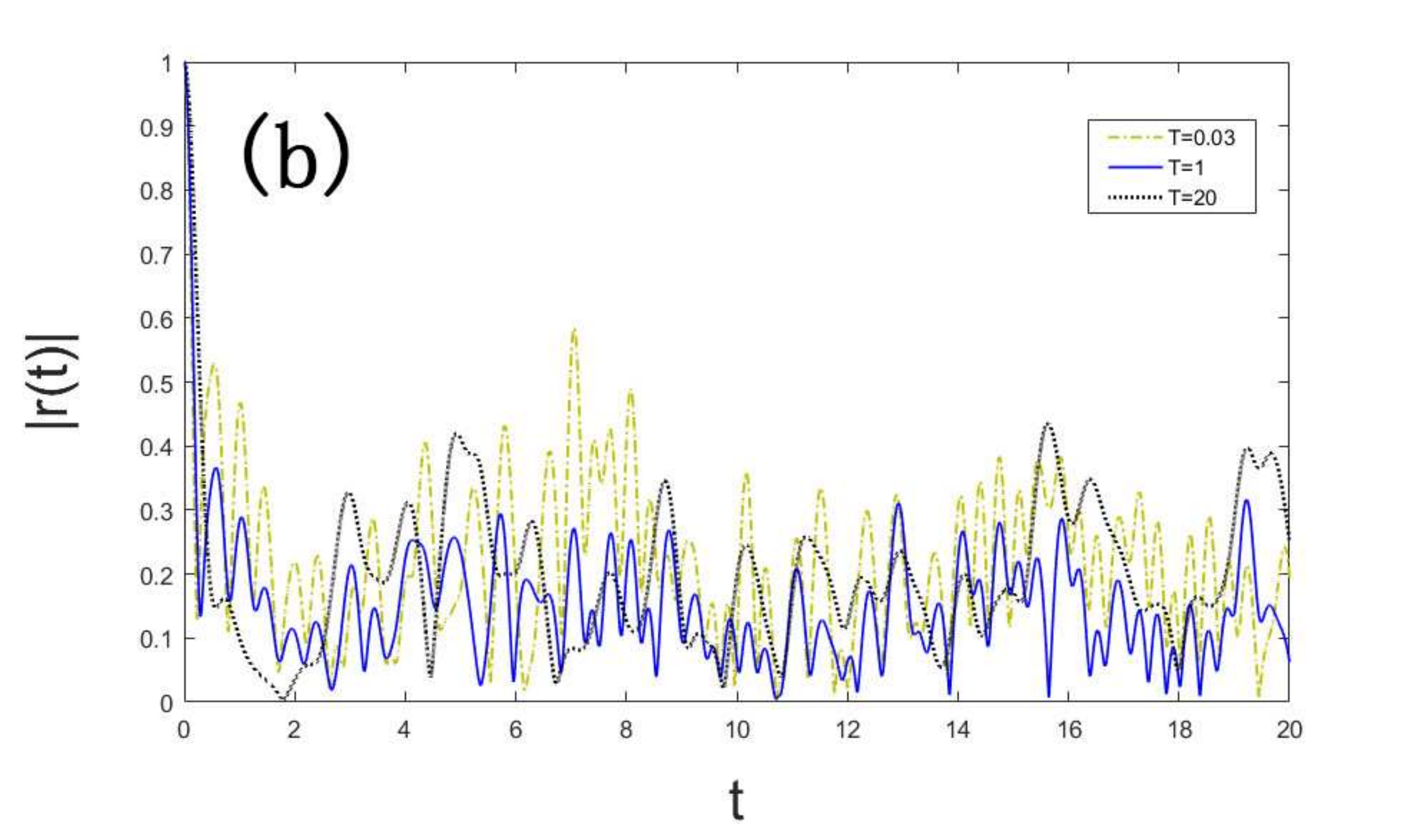}
  }
  \caption{Real-time evolutions of the coherence factor $|r(t)|$ of the central spin in the antiferromagnetic bath (a) and the ferromagnetic
bath (b) at different temperatures. In both plots, we use $g=1$, $\omega=10$, $\Delta=5$, and $N=8$.}
\end{figure}

We study the antiferromagnetic environment and the ferromagnetic environment separately to identify the different evolution behaviors of the central spin coherence in these two states.
The coherence factors $|r(t)|$ for the bath spins with antiferromagnetic coupling and ferromagnetic coupling at different temperatures are shown in Figs.\,\ref{fig4:subfig:a} and \ref{fig4:subfig:b}, respectively.
There is no obvious lower bound for the case with a ferromagnetic bath where the oscillation frequency of the coherence is high. The lower bound of $|r(t)|$ for the case with an antiferromagnetic bath decreases with increasing temperature $T$ due to the fact that more Bethe states are occupied by the bath spins and involved in the evolution of the central spin at high temperatures. When the temperature tends to infinity, the weight factors $e^{-E_x/(k_BT)}$ of all bath eigenstates are about the same for both antiferromagnetic and ferromagnetic baths. Hence the evolution curves with high temperatures are similar and the coherence of the central spin becomes poor.

\begin{figure}[htbp]
  \centering
  \subfigure{
    \label{fig5:subfig:a} 
    \includegraphics[width=3.4in, height=2.6in]{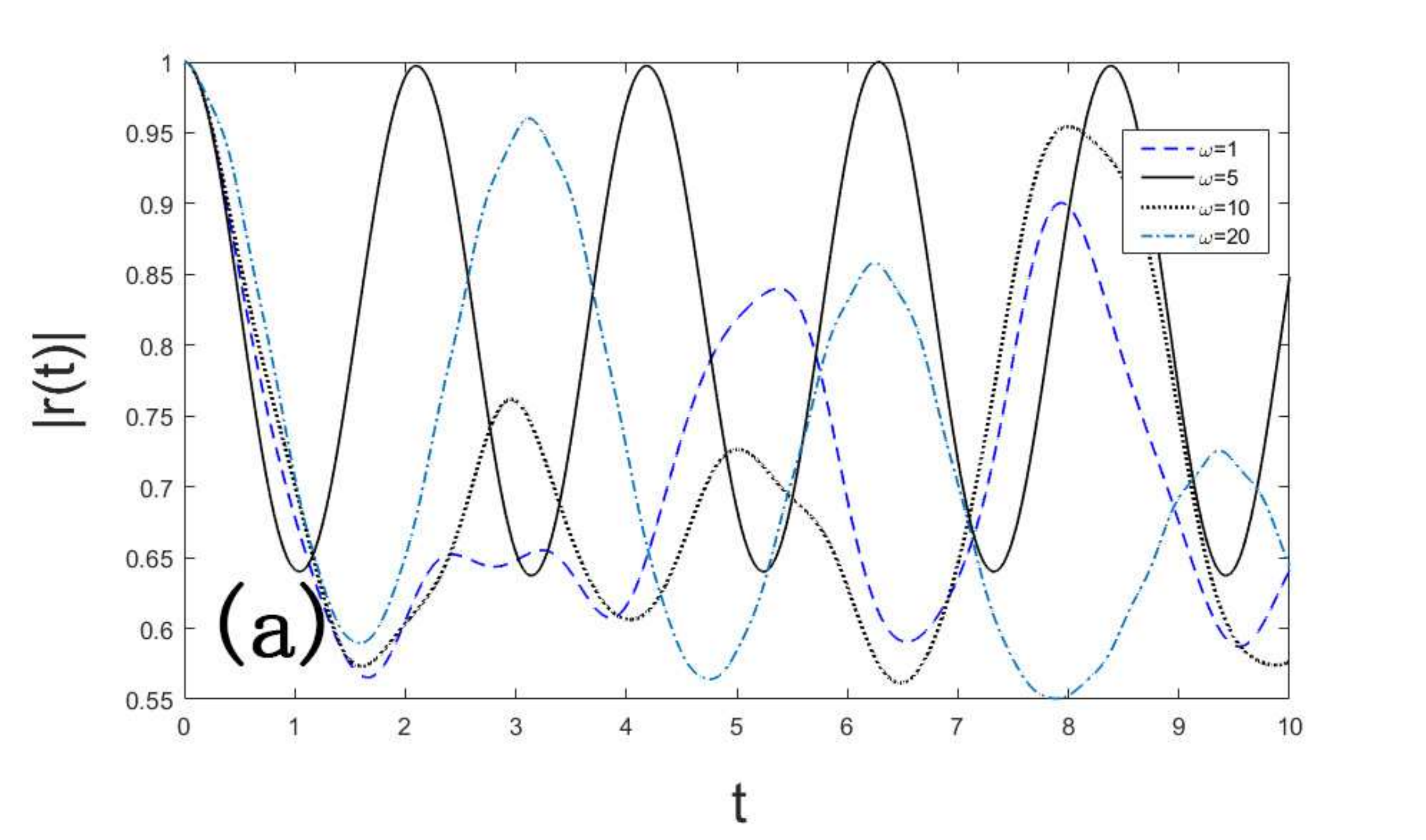}
  }
  \hfill
  \subfigure{
    \label{fig5:subfig:b} 
    \includegraphics[width=3.4in, height=2.6in]{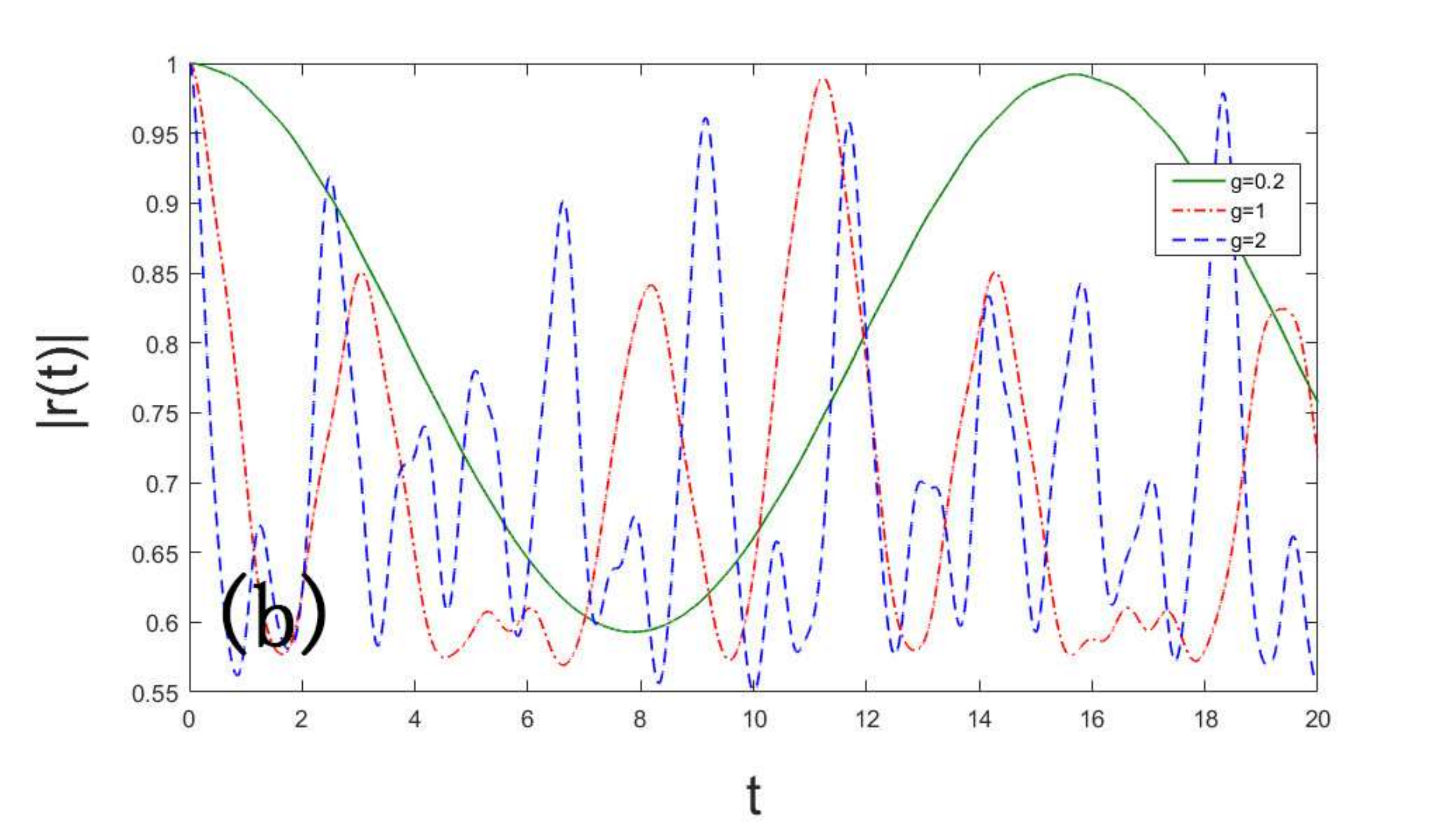}
  }
  \caption{Real-time evolutions of the coherence factor $|r(t)|$ of the central spin in the antiferromagnetic environment at finite temperature for varying $\omega$ but $g=1$ fixed (a) and for varying $g$ but $\omega=12$ fixed (b).  In both plots, we use $T=1$, $\Delta=5$, and $N=8$.}
\end{figure}

At low temperatures, the Bethe states $|\lambda_1,\cdots, \lambda_M; S_M\rangle$ with small $S_M$ dominate for the antiferromagnetic coupling,
while the Bethe states with large $S_M$ dominate for the ferromagnetic coupling, which results in the obvious difference between the coherence behaviors with an antiferromagnetic bath and a ferromagnetic one. For the antiferromagnetic bath, the lower the temperature is, the better the central spin coherence is.
We note that the number of possible states is larger with larger $S_M$ and the degeneracy of Bethe states is higher. Since more states are involved in the evolution for the case with ferromagnetic coupling, the central spin coherence is not as good as that with antiferromagnetic coupling at low temperatures.

The evolutions of the coherence factor $|r(t)|$ in the antiferromagnetic environment at finite temperatures with fixed $g$ and fixed $\omega$ are shown in Figs.\,\ref{fig5:subfig:a} and \ref{fig5:subfig:b}, respectively, while keeping $\Delta=5$ and $N=8$. As shown in Fig.\,\ref{fig5:subfig:a}, there is a critical external magnetic field of the bath spins whose value is the same as the external magnetic field of the central spin, i.e., $\omega=\Delta$, which gives nearly perfect periodical oscillation of the coherence. This behavior is completely contrary to the coherence evolution when the bath spins occupy their eigenstates, where the lower bound is the smallest at the critical $\omega_c=12$ as shown in Fig.2(a). Meanwhile, the coherence will not improve with increasing $\omega$, which is also contrary to the result in Fig.2(a). This is because when we consider finite temperatures, all the possible states are included.

Let us give some detailed analyses at the critical point $\omega=\Delta$. From Eq.\,(\ref{eq3.21}) and considering the fixed $g$, at the critical point of $\omega_c^{\prime}=\Delta$, we have
\begin{eqnarray}
\varepsilon_n^2+g_n^2=\zeta_n^2=g^2(S_M^2+S_M+1/4),\hspace{0.5cm} n\in[-S_M, S_M],\label{eq5.6}
\end{eqnarray}
with $\zeta_n$ related to the evolution frequency of the coherence as shown in Eq.\,(\ref{eq5.4}). This indicates that $\zeta_n$ with different quantum numbers $n$ but the same highest-weight $S_M$ are equal, which can be regarded as that the number of permitted occupation states being reduced.

\begin{figure}[htbp]
\centering
\includegraphics[scale=0.5]{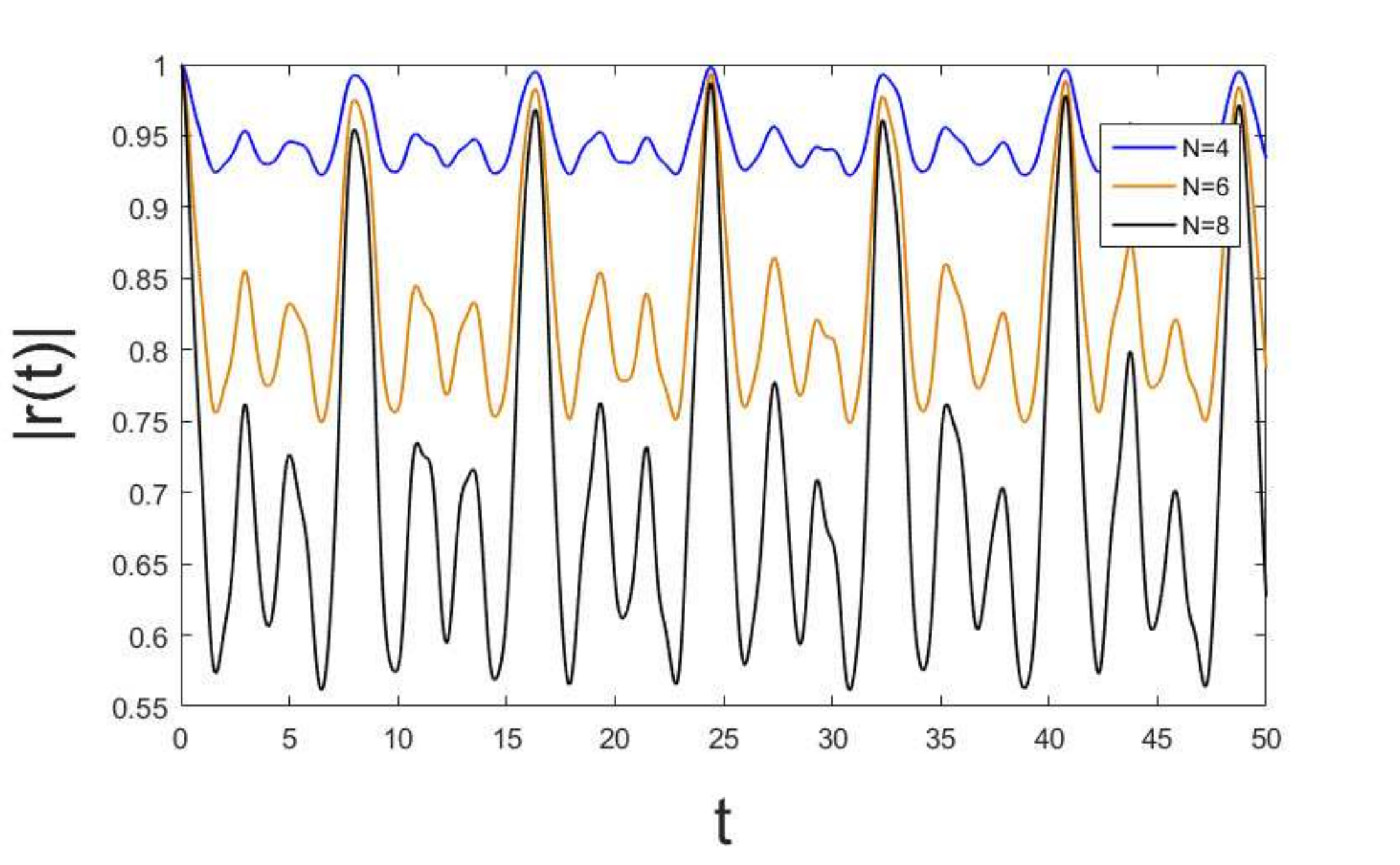}
  \caption{Real-time evolutions of the coherence factor $|r(t)|$ of the central spin in the antiferromagnetic environment at finite temperature with different numbers of bath spins, where $g=1$, $\omega=10$, $\Delta=5$, and $T=1$.}\label{fig6}
\end{figure}

With $\omega$ fixed in Fig.\,\ref{fig5:subfig:b}, the decoherence is quick at the beginning of the evolution of the system with increasing $g$. For large $g$, the oscillation decoherence is obvious. Due to the facts that $S_M$ is large at low temperatures and that we do the summation of all possible states, the coupling between the central spin and the bath spins does not have a critical value. Moreover, the evolution of the coherence is affected by the number of bath spins. The larger the number of bath spins $N$ is, the more states that are involved in the evolution. As shown in Fig.\,\ref{fig6}, reducing the number of bath spins suppresses the decoherence of the central spin greatly.

\section{Evolution of the central spin polarization}
\label{sec5} 

In this section, we consider the dynamical properties of the central spin polarization.
The time-dependent spin polarization along the $z$ direction is defined as
\begin{eqnarray}\label{eq6.1}
S_0^z(t)=tr\{S_0^z U(t) \rho_{tot}(0) U^{\dagger}(t)\}=tr\{S^z \rho_{c}(t)\},
\end{eqnarray}
by using the central spin operator along the $z$-direction, $S^z$, and the
reduced density matrix of the central spin at the time $\rho_c(t)$,
The spin polarization at the initial bath thermal states reads
\begin{eqnarray}\label{eq6.2}
S_0^z(t)=\frac{1}{Z}\sum_{M=0}^{[\frac{N}{2}]}\sum_{\{\lambda_1, \cdots, \lambda_M\}}\sum_{n=-S_M}^{S_M}e^{-\frac{E_x(\lambda_1,\cdots, \lambda_M)}{k_BT}}\rho^{11}_{c, M, n}(t),
\end{eqnarray}
where $\rho^{11}_{c, M, m}(t)$ is the element of $\rho_c(t)$ given by Eqs.\,(\ref{eq4.4})-(\ref{eq4.7}).
Clearly, the value of $S_0^z(t)$ depends on the combination coefficients $\alpha$ and $\beta$ of the initial state of the central spin. We take the parametrizations $\alpha=\sin\theta$ and $\beta=\cos\theta$ for simplicity. The central spin polarization $S_0^z(0)=(\sin^2\theta-\cos^2\theta)/2$ at $t=0$. For the given value of $\theta=0.4\pi$, the spin polarization starts from $0.40451$. The same as before, we consider the antiferromagnetic bath and the ferromagnetic bath separately to discover the different effects on the central spin polarization.

\begin{figure}[htbp]
  \centering
  \subfigure{
    \label{fig7:subfig:a} 
    \includegraphics[width=3.3in, height=2.5in]{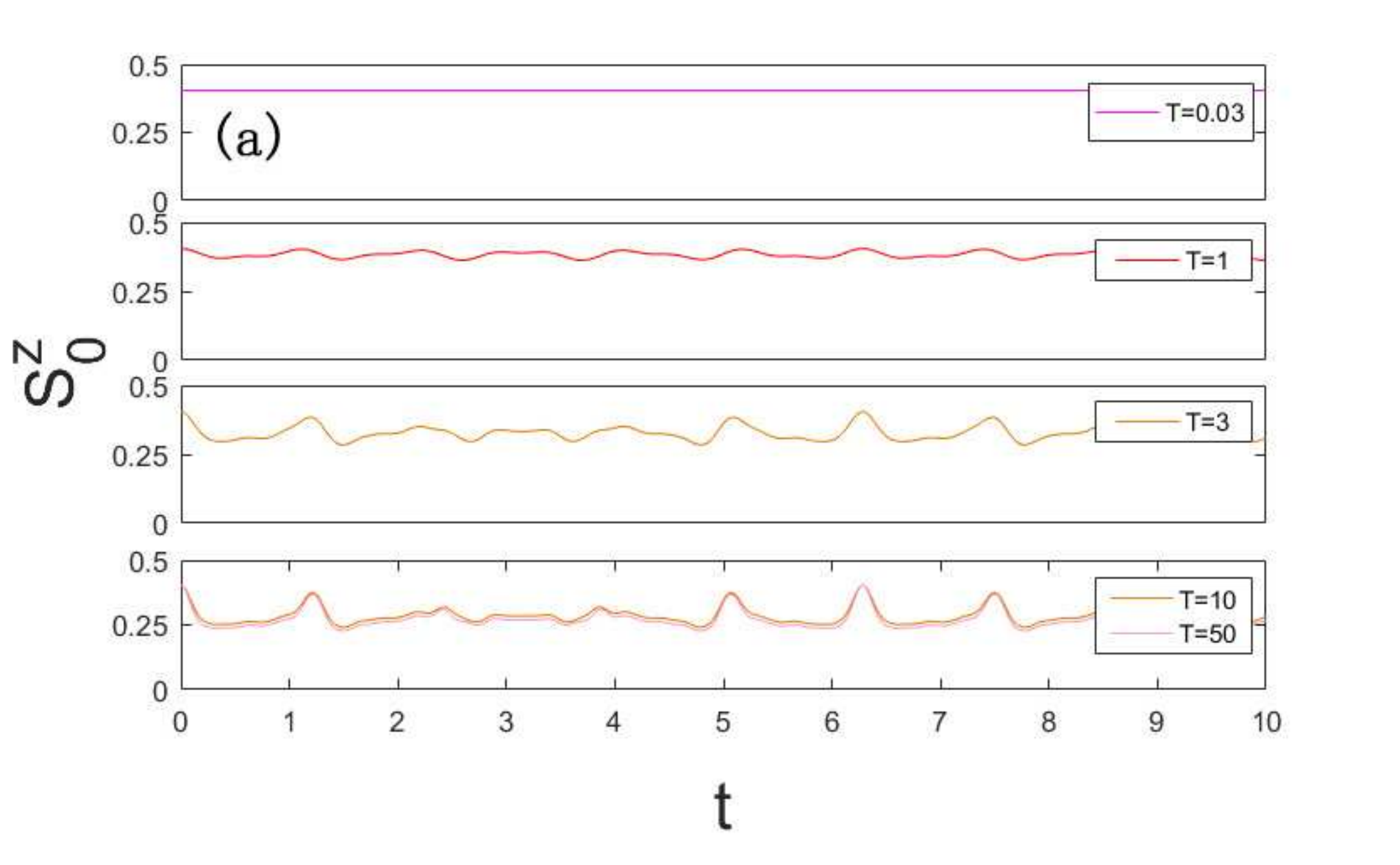}
  }
  \hfill
  \subfigure{
    \label{fig7:subfig:b} 
    \includegraphics[width=3.3in, height=2.5in]{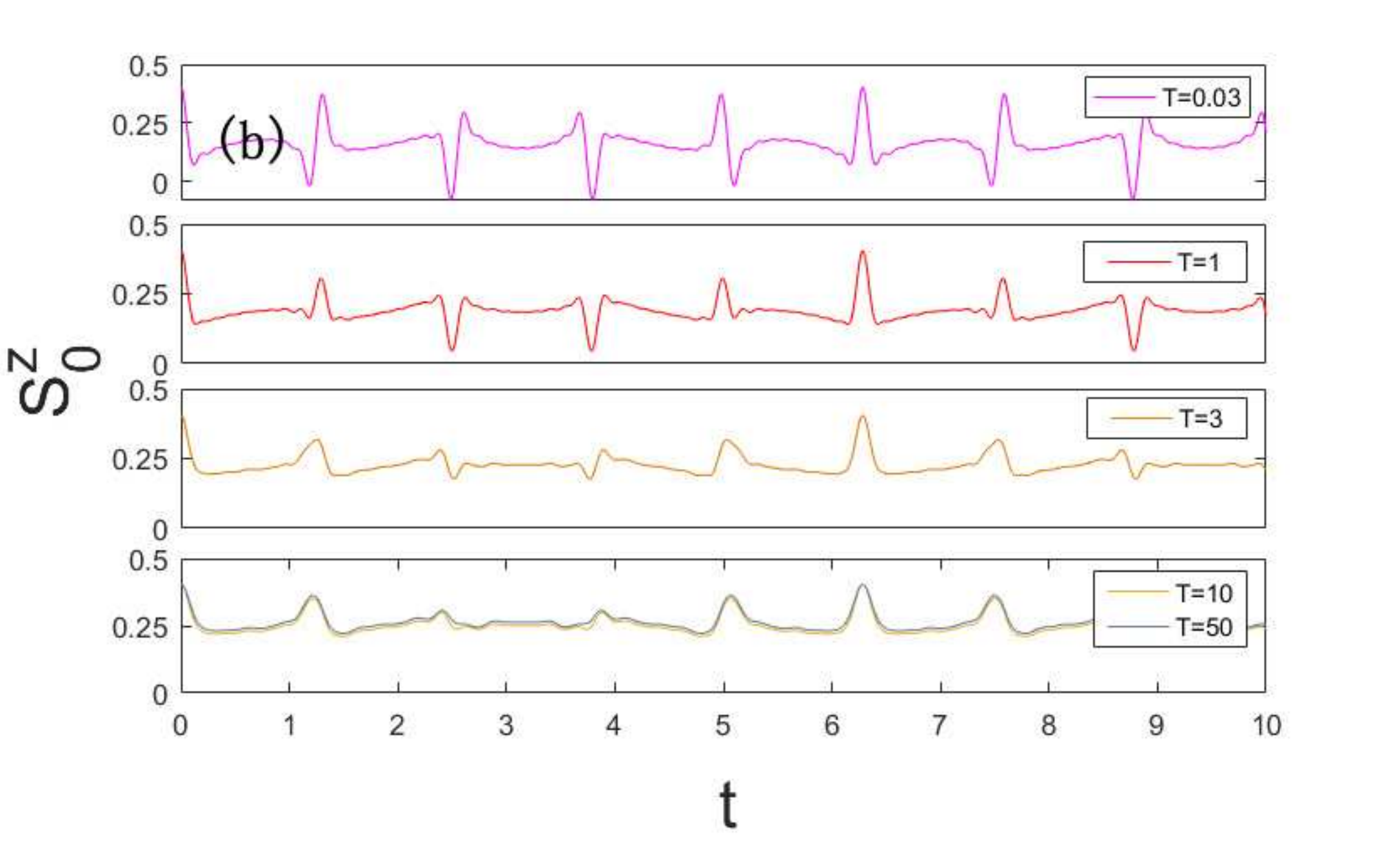}
  }
  \caption{Real-time evolutions of the spin polarization $S_0^z(t)$ of the central spin at different temperatures in the antiferromagnetic bath (a) and the ferromagnetic bath (b). In both plots, we use $g=1$, $\omega=10$, $\Delta=5$, $\theta=0.4\pi$, and $N=8$.}
\end{figure}

In Fig.\,\ref{fig7:subfig:a}, we show the time evolution of the spin polarizations at different temperatures in the antiferromagnetic bath. We see that the spin polarization is suppressed with the increase of the temperature. At zero temperature, the antiferromagnetic bath spins occupy their ground state $|\lambda_1,\cdots, \lambda_M; S_M\rangle$, with $S_M=0$.
The initial state is $|\lambda_1,\cdots, \lambda_M; S_M\rangle \otimes (\sin\theta |\uparrow\rangle+\cos\theta|\downarrow\rangle)$.
We have proved that the states $|\lambda_1,\cdots, \lambda_M; S_M\rangle\otimes|\uparrow\rangle$ and $|\lambda_1,\cdots, \lambda_M; -S_M\rangle\otimes|\downarrow\rangle$
are the eigenstates of the system (\ref{eq2.2}) in Sec.\,\ref{sec3.1}, which means $S_0^z(t)$ does not evolve with time. With increasing temperature, the weight of this state decreases, which makes the spin polarization decreases as well. For the ferromagnetic bath, the spin polarization oscillates stronger at lower temperatures, while the mean value increases with the temperature as shown in Fig.\,\ref{fig7:subfig:b}. These behaviors are contrary to those for the antiferromagnetic bath shown in Fig.\,\ref{fig7:subfig:a}.

\begin{figure}[htbp]
  \centering
  \subfigure{
    \label{fig8:subfig:a} 
    \includegraphics[width=3.3in, height=2.5in]{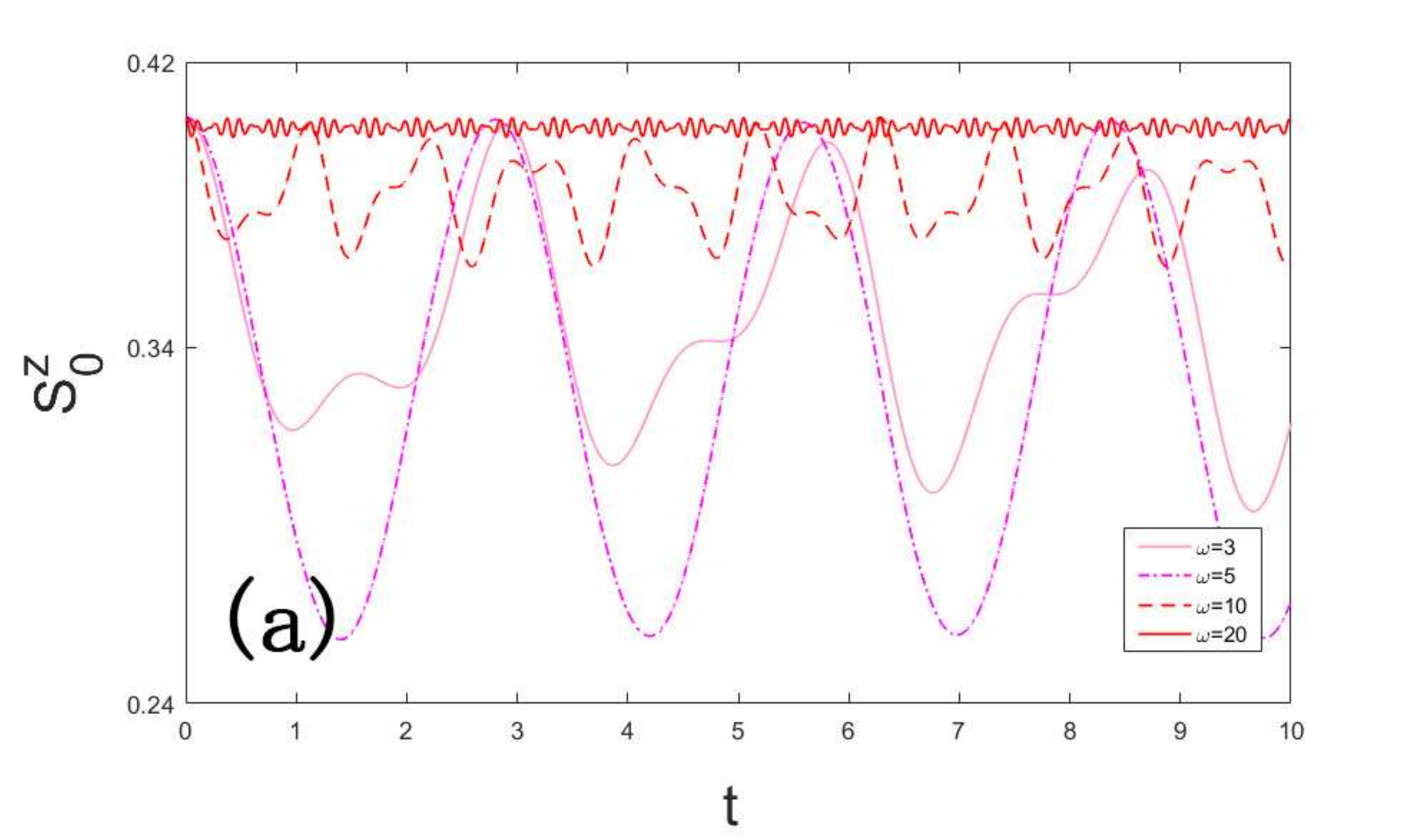}
  }
  \hfill
  \subfigure{
    \label{fig8:subfig:b} 
    \includegraphics[width=3.3in, height=2.5in]{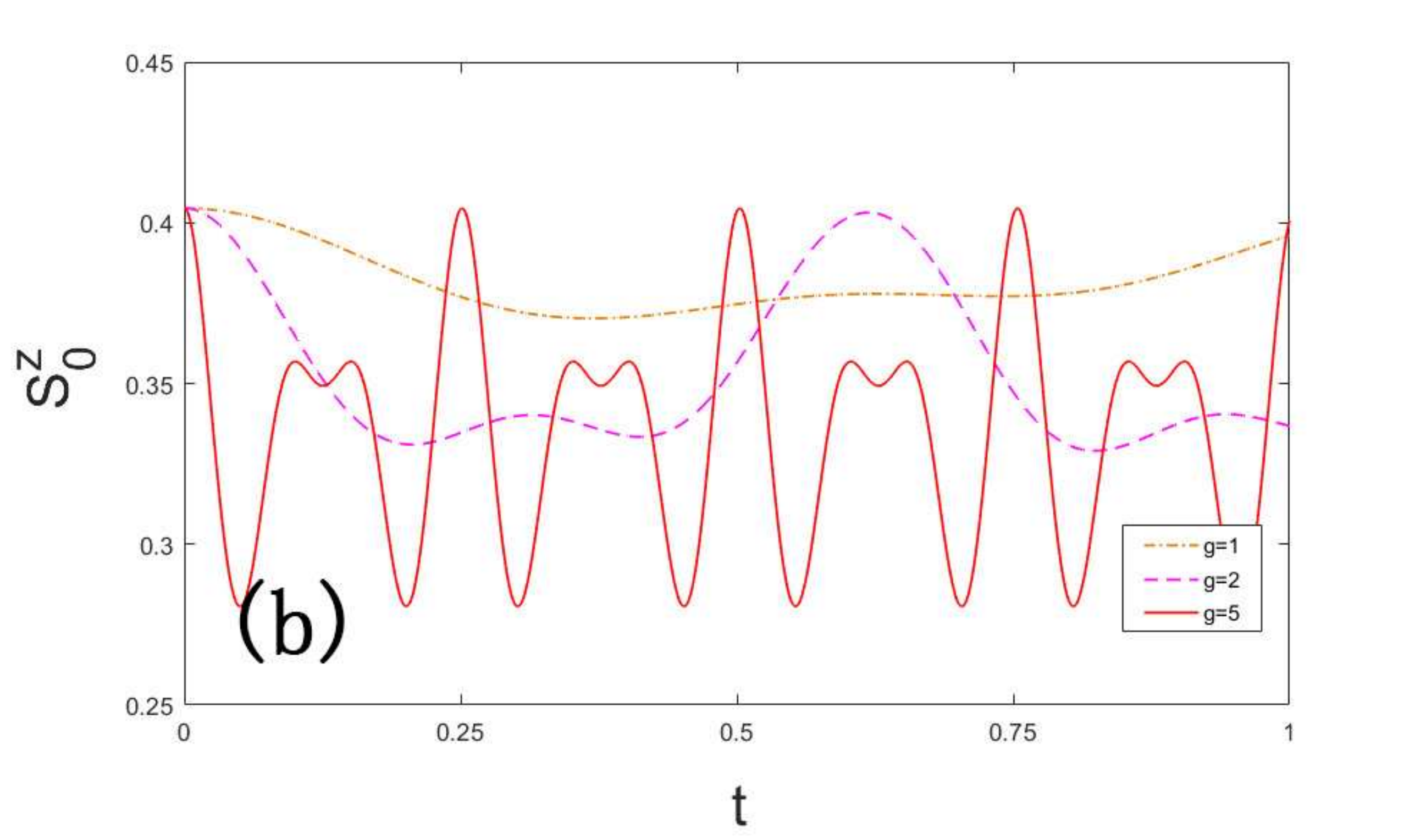}
  }
  \caption{Real-time evolutions of the spin polarization $S_0^z(t)$ of the central spin in the antiferromagnetic environment at finite temperature for varying $\omega$ but $g=1$ (a) and for varying $g$ but $\omega=50$ (b). In both plots, we use $T=1$, $\Delta=5$, $\theta=0.4\pi$, and $N=8$.}
\end{figure}

An interesting dephasing process is discovered, which means that the diagonal elements of the reduced density matrix remain unchanged while the non-diagonal elements evolve. This phenomenon appears at $\theta={\pi/4}$, which means that the weights of the $|\uparrow\rangle$ and $|\downarrow\rangle$ states are the same. The initial spin polarization is zero according to $(\langle |\uparrow|+\langle \downarrow|)S^z(|\uparrow\rangle+|\downarrow\rangle)=0$. Through detailed calculations, we find that $S_0^z(t)$ does not evolve for any temperatures and other system parameters.
However, the result does not suggest that the central spin keeps the initial coherence properties. Because the central spin coherence will evolve for other initial values of $\theta$ as stated in Sec.\,\ref{sec4}, the central spin and the bath spins do exchange information.

The evolution of the spin polarization also changes with $\omega$ and $g$.
For fixed $g$, we find that the evolution of spin polarization has an upper bound and a lower bound as shown in Fig.\,\ref{fig8:subfig:a}, where the upper bound is nothing but the initial spin polarization. The lower bound and the oscillation frequency increase with increasing $|\omega-\omega^{\prime}_c|$, where $\omega^{\prime}_c=\Delta$ is the critical point. The information exchange between the central spin and the bath spins is enhanced at the critical point $\omega=\omega^{\prime}_c$ as shown by {the mauve dashed line in Fig.\,\ref{fig8:subfig:a}}.
As shown in Fig.\,\ref{fig8:subfig:b}, the lower bound of the spin polarization with fixed $\omega$ decreases while the oscillation frequency increases with increasing $g$. This means that the information exchange becomes more frequent with stronger couplings between the central spin and the bath spins. For the evolution of the central spin polarization with varying $\omega$ and $g$ in the ferromagnetic bath, the system has properties similar to those in the antiferromagnetic bath.

\section{Conclusions}
\label{sec6} 

In this paper, we have studied the extended central spin model with XXX isotropic spin-exchanging interaction between the bath spins. By using the Bethe states of the bath spins and the spin-flipping operators, we construct a complete basis of the Hilbert space for the central spin system.
We find that the Hilbert space can be divided into the direct product of some two-dimensional invariant subspaces. Thus
the Hamiltonian can be expressed by the direct sum of certain 2$\times$2 matrix blocks.
By solving the eigen-equations, we obtain the exact eigenstates and the associated eigenvalues of the extended central spin system.

With the help of the exact solutions, we obtain the reduced density matrix and study the coherence of the central spin in different situations.
If the bath spins occupy an eigenstate, we find that the external magnetic fields ($\omega$ and $\Delta$), the couplings between the central spin and the bath spins along
the $z$ direction, and the magnetization of the bath spins (measured by the quantum number $m$) can constitute a total magnetic field $\varepsilon$ applied on
the central spin. Thus the central spin coherence can be enhanced by increasing $|\varepsilon-\varepsilon_c|$ or by decreasing
the coupling between the central spin and the bath spins. With the initial thermal state of the bath spins, the central spin coherence in the antiferromagnetic environment can be enhanced by decreasing the temperature or the number of the bath spins.
We also find that the central spin has good coherence at the critical point of $\omega=\Delta$.
The decoherence rate in an initial short time is slower if the coupling is weak.

The dynamical behaviors of the central spin polarization has been studied under different circumstances.  With increasing temperature, the spin polarization decreases in the antiferromagnetic bath while it increases in the antiferromagnetic bath. A dephasing process is discovered in our central spin model, where the central spin polarization is fixed even though there is information exchange between the central spin and the bath spins. Moreover, the information exchange between the central spin and the bath spins is enhanced at $\omega=\Delta$, which induces stronger oscillation of the central spin polarization. At low temperatures, the central spin coherence and the polarization in the antiferromagnetic bath and ferromagnetic bath show completely different evolution properties. The antiferromagnetic bath gives much better coherence and polarization of the central spin. These results are helpful for further insight into the mechanisms of decoherence and are useful in selecting optimal configurations from the enormous freedom of the solid states in real implementation. We also note that our method can be applied to study the extended central spin models with bath spins in an isotropic nearest-neighbour interaction  spin chain with a higher spin or spin chains associated with higher-rank algebras such as $su(n)$, $so(n)$, $sp(2n)$, etc. The generalization to a system with an anisotropic bath spin interaction (such as the XXZ or XYZ spin chain) is still an interesting open problem.


\section*{Acknowledgments}

We thank to thank Professor Y. Wang for his valuable suggestions and continuous encouragement.
Financial support from the National Program
for Basic Research of MOST (Grants No. 2016YFA0300600 and
No. 2016YFA0302104), the National Natural Science Foundation of China
(Grants No. 11975183, No. 11934015, No. 11774397, No. 11775178, No. 11775177, and No. 11947301), the Major Basic Research Program of Natural Science of Shaanxi Province
(Grants No. 2017KCT-12, and No. 2017ZDJC-32), the Australian Research Council (Grant No. DP 190101529), the Strategic Priority Research Program of the Chinese
Academy of Sciences, and the Double First-Class University Construction Project of Northwest University is gratefully acknowledged.
P.L. thanks K. Hao, P. Sun, and F. Wen for helpful discussions.


%

\end{document}